\documentclass[a4paper,usenames,dvipsnames,11pt]{article}

\usepackage{jheppub}
\usepackage{slashed}
\usepackage{mathrsfs,booktabs}
\usepackage{stmaryrd}
\usepackage{xspace}
\usepackage{fancyvrb}
\usepackage{multirow}
\usepackage{pifont}
\usepackage{scalefnt}
\usepackage{float}

\newcommand{\eq}[1]{eq.~\eqref{eq:#1}}
\newcommand{\eqs}[2]{eqs.~\eqref{eq:#1} and \eqref{eq:#2}}
\renewcommand{\sec}[1]{section~\ref{sec:#1}}

\newcommand{\subsec}[1]{section~\ref{subsec:#1}}

\newcommand{\fig}[1]{figure~\ref{fig:#1}}

\newcommand{\mycite}[1]{ref.~\cite{#1}}
\newcommand{\mycites}[1]{refs.~\cite{#1}}
\newcommand{\tab}[1]{table~\ref{tab:#1}}

\newcommand{\gamt}{\Gamma_t}
\newcommand{\gamh}{\Gamma_H}
\newcommand{\as}{\alpha_s}
\newcommand{\muR}{\mu_R}
\newcommand{\muF}{\mu_F}
\newcommand{\mt}{m_t}
\newcommand{\rjet}{R_{\text{jet}}}

\allowdisplaybreaks

\newcommand{\abbrev}{\rm\scalefont{.9}}

\newcommand{\sm}{{\abbrev SM}}
\newcommand{\bsm}{{\abbrev BSM}}
\newcommand{\lhc}{{\abbrev LHC}}

\newcommand{\mga}{{\sc{MadGraph5\_aMC@NLO}}\xspace}
\newcommand{\whz}{{\sc{WHIZARD Event Generator}}\xspace}

\numberwithin{equation}{section}
\VerbatimFootnotes

\title{Probing the top-quark width through ratios of resonance contributions of $e^+e^-\rightarrow W^+W^-b\bar{b}$}

\author[a]{Stefan Liebler,}
\author[ab]{Gudrid Moortgat-Pick}
\author[ac]{and Andrew S. Papanastasiou}
\affiliation[a]{DESY, Deutsches Elektronen-Synchrotron,\\ Notkestra{\ss}e 85,D-22607 Hamburg, Germany}
\affiliation[b]{II. Institut f{\"u}r Theoretische Physik, Universit{\"a}t Hamburg,\\ Luruper Chaussee 149, 22761 Hamburg, Germany}
\affiliation[c]{Cavendish Laboratory, University of Cambridge,\\ J.J. Thomson Avenue, CB3 0HE, Cambridge, UK}

\emailAdd{stefan.liebler@desy.de}
\emailAdd{gudrid.moortgat-pick@desy.de}
\emailAdd{andrewp@hep.phy.cam.ac.uk}

\abstract{
We exploit offshell regions in the process $e^+e^-\rightarrow W^+W^-b\bar{b}$ to gain access to the top-quark width.  
Working at next-to-leading order in QCD we show that carefully selected ratios of offshell regions to onshell regions 
in the reconstructed top and antitop invariant mass spectra are, \emph{independently} 
of the coupling $g_{tbW}$, sensitive to the top-quark width.
We explore this approach for different centre of mass energies and initial-state beam polarisations at $e^+e^-$ colliders and 
briefly comment on the applicability of this method for a measurement of the top-quark width at the LHC.
} 


\preprint{
\begin{flushright}
Cavendish-HEP-15/11 \\
DESY 15-169 \\
\today
\end{flushright}
}

\begin{document}

\maketitle
\flushbottom

\newpage

%%%%%%%%%%%%%%%%%%%%%%%%%%%%%%%%%%%%%%%%%%%%%%%%%%%%%%%%%%%%%%%%%%%%%%%%%%%%%%%%%%%%%
\section{Introduction} \label{sec:intro}
%%%%%%%%%%%%%%%%%%%%%%%%%%%%%%%%%%%%%%%%%%%%%%%%%%%%%%%%%%%%%%%%%%%%%%%%%%%%%%%%%%%%%

After the spectacular discovery of a signal in the Higgs searches
at the Large Hadron Collider (\lhc{})~\cite{Aad:2012tfa,Chatrchyan:2012xdj}
a future linear collider,
such as the International Linear Collider
~\cite{Behnke:2013xla,Baer:2013cma,Adolphsen:2013jya,Adolphsen:2013kya,Behnke:2013lya},
provides rich and exciting physics prospects in the context
of the Standard Model (\sm) of particle physics and of course beyond.
Aside from the detailed properties of the Higgs boson,
which can be measured with high accuracy at a linear collider, the precise determination
of top-quark properties is also of high priority, as these may also provide an interesting window to new physics.
For recent reviews of linear collider physics we refer to \mycites{Moortgat-Pick:2015yla,Fujii:2015jha}
and a comprehensive review on top-quark physics at a linear collider can be found within \mycites{Juste:2006sv,Agashe:2013hma}.

It is well-known that the precise measurement of many top-quark properties, including that
of a well-defined mass parameter, is optimally performed by studying the production of
a top-quark pair near threshold, that is at a centre of mass $\sqrt{s}\sim 2\mt\sim 350$\,GeV
where the two top quarks are produced almost at rest and show strong binding effects.
However, it is not the purpose of this paper to revisit this phase space region and we 
refer to \mycites{Moortgat-Pick:2015yla,Fujii:2015jha} and the plethora of references therein
for details and insights. 
A linear collider is likely to initially operate at a higher centre of mass energy
$\sqrt{s}\geq 500$\,GeV~\cite{Barklow:2015tja},
where top-quark pairs are produced in the continuum rather than at rest.
There are a number of good reasons to start a linear collider run at a higher centre of mass 
energy including: making a measurement of the top-quark Yukawa coupling possible early on, 
recording of Higgsstrahlung and vector-boson fusion events and thus determining the couplings 
of the Higgs boson to heavy gauge bosons and, of course, better probing regions of phase
space where new physics is more likely to appear. 
Therefore, despite the fact that a run at threshold will allow
the extraction of theoretically more consistent and experimentally easier to measure
top-quark properties, it is important to explore what possibilities 
for precise measurements of top-quark properties are offered by the continuum region.

Top-quark production in the continuum in $e^+e^-$ collisions has and continues to be 
a subject of much theoretical attention. 
Onshell top-quark pair-production in the continuum is now known at next-to-next-to-leading order (NNLO) in QCD 
\cite{Chetyrkin:1997mb,Hoang:1997ca,Harlander:1997kw,Chetyrkin:1997pn,Chetyrkin:2000zk,
Bernreuther:2004ih,Bernreuther:2004th,Bernreuther:2005rw,Gluza:2009yy,Gao:2014nva,Gao:2014eea}.
Relaxing the assumption of stable top quarks, top-pair production and decay in $e^+e^-$ collisions has been 
studied in the narrow-width approximation in \mycite{Schmidt:1995mr}. 
The full process $e^+e^- \to W^+W^-b\bar{b}$, in which intermediate top quarks can have arbitrary offshellness
was first computed at next-to-leading order (NLO) in \mycite{Lei:2008ii} and recently revisited in \mycites{Mattelaer:2014ewa,Weiss:2015npa}.
Leading order (LO) predictions for $e^+e^- \to 6$~fermions can be found in \cite{Dittmaier:2002ap}.
A discussion of top-quark production with unstable top quarks in the continuum and in the regime of boosted tops,
using a tower of effective field theories was provided in \mycites{Fleming:2007qr,Fleming:2007xt}. The latter
compute the double differential cross section with respect to the invariant masses of the two top quarks
to next-to-leading-log, and show that in such a regime a well-defined mass parameter can, in principle, be determined with 
an accuracy of better than $\mathcal{O}(\Lambda_{\text{QCD}})$.

In this work we explore the idea of obtaining the top-quark width, $\gamt$,
by exploiting the different resonance regions in the reconstructed top and antitop invariant 
mass distributions that are present in the process $e^+e^- \to W^+W^-b\bar{b}$.
Top-pair and single top production co-exist in this full process and contributions to 
the different regions can be identified as double-resonant or single-resonant top-quark 
production, which intrinsically differ in their dependence on $\gamt$.
For our investigation we simulate the fully-differential process using \mga~\cite{Alwall:2014hca} at LO and 
NLO in QCD and emphasise that the full set of diagrams for $e^+e^- \to W^+W^-b\bar{b}$ (i.e. those
with two, one and no intermediate top-quark propagators) is included in the calculation. 

Our paper is organized as follows:
In the remainder of \sec{intro}, we motivate our approach by showing an analogy to the determination
of the Higgs boson width from offshell regions and later transfer 
the idea to the case of the top-quark width, 
for which the current theoretical and experimental knowledge
is also briefly reviewed. In \sec{proc-defn} we discuss the
process $e^+e^- \to W^+W^-b\bar{b}$ at NLO QCD, including our numerical
setup and details on relevant distributions. We also provide a detailed 
examination of the reconstructed top-quark mass distribution, thus gaining insight 
into the structure of the different resonance regions of this process.
In \sec{offshell-width} we show how the method for the Higgs boson is
modified in the case of a pair of unstable top quarks. We apply this method to
$e^+e^- \to W^+W^-b\bar{b}$ and illustrate how it enables one to gain access to the top-quark width.
We additionally investigate the potential for enhanced sensitivities by exploiting 
polarised beams or higher centre of mass energies.
Finally, we also discuss how our analysis may be improved in future studies and comment on
the applicability of our method at the LHC. We end with our conclusions in \sec{conclusion}.

\subsection{Offshell regions and the Higgs boson width} \label{subsec:gamH-offshell}

After the discovery of a \sm{}-like Higgs boson at $125$\,GeV,
offshell contributions in the decay of the \sm{} Higgs into a pair of vector bosons ($V$)
were found to be sizeable~\cite{Kauer:2012hd,Kauer:2013cga,Kauer:2013qba}.
Offshell region measurements have offered an opportunity to indirectly constrain the width of the Higgs boson
at the LHC via the method proposed in \mycites{Caola:2013yja,Campbell:2013una,Campbell:2013wga}.
The key idea is that the ratio of offshell to onshell cross section
measurements is sensitive to the total Higgs width, $\gamh$. This can be inferred by examining how the 
cross sections in the different regions scale with the couplings involved in Higgs production and decay 
and how they scale with $\gamh$. The onshell cross section receives contributions
from phase space where the invariant mass of the vector-boson pair is close to the Higgs boson pole mass, $M(V,V) \sim m_H$, 
and scales as $\sigma^{\text{on}}_{VV} \sim g^2_{\text{on}} \gamh^{-1}$, where $g_{\text{on}}$ encodes
all couplings involved in Higgs production and decay. 
In contrast, events with $M(V,V) \gg m_H$ contribute to 
the offshell cross section, which scales as $g^2_{\text{off}}$, i.e. it is independent of the Higgs boson width~$\gamh$. 

Under the assumption that the couplings in the offshell region $g_{\text{off}}$ can be related to those in the 
onshell region $g_{\text{on}}$ then an extraction of the width is possible simply by relating
the on- and offshell signal strength. For details in case of $e^+e^-$ collisions
we refer to \mycite{Liebler:2015aka}, where
offshell effects in Higgs production at a linear collider were discussed.
It must be emphasised that the extracted bound on the Higgs width ought to be taken with some 
care as it is based on an assumed relation between on- and offshell couplings,
namely that the onshell and offshell $\kappa$ factors are equal.\footnote{For a definition
of the $\kappa$-factors framework, we refer to \mycites{Heinemeyer:2013tqa,LHCHiggsCrossSectionWorkingGroup:2012nn}.}
The latter relation can be severely affected by Beyond-the-Standard Model (\bsm{}) physics
as discussed in~\mycites{Ghezzi:2014qpa,Englert:2014aca,Englert:2014ffa,Logan:2014ppa}
for the \lhc{} and in \mycite{Liebler:2015aka} for a linear collider.

\subsection{Offshell regions and the top-quark width} \label{subsec:gamt-offshell}
In \sec{proc-defn} and \sec{offshell-width} we investigate whether a similar procedure of
relating cross sections in different kinematic regimes can be applied to the case of top-quark production
to infer the total width of the top quark, $\gamt$. For this purpose we consider the
process $e^+e^-\rightarrow W^+W^-b\bar{b}$. This is the relevant final state for the top-quark pair production process
where the decay of the tops is included. However, the full perturbative calculation we work with contains single resonant
and non resonant contributions in addition to the usual double-resonant contributions and also includes full finite-top-width effects.
We consider the various resonance regions present in the double differential cross section
%%%
\begin{align}
\frac{d^2\sigma^{e^+e^- \to W^+W^-b\bar{b}}}{d M(W^+,J_b) \, d M(W^-,J_{\bar{b}})}
\,,\end{align}
%%%
where $M(W^+,J_b)$ and $M(W^-,J_{\bar{b}})$ are the top and antitop masses reconstructed through 
the $W$-bosons and $b/\bar{b}$-flavoured jets, $J_b/J_{\bar{b}}$, present in the final state.
The different resonance regions  
are influenced to varying degrees by all of the double resonant (`top-pair'),
single resonant (`single top') and non resonant (`no top') subprocesses
to $e^+e^- \to W^+W^-b\bar{b}$.

There are a few differences between the method used to place bounds on the Higgs boson width and the 
one we propose here to become sensitive to $\gamt$.
Firstly, unlike in the case of the Higgs boson, we work in a limited kinematic range rather 
close to the top quark resonance peaks $M(W^+,J_b),\; M(W^-,J_{\bar{b}})\sim \mt\pm 50$\,GeV.
In this range the influence of possible `high-mass' \bsm{} contributions
is therefore limited and we can safely treat the involved couplings, most prominently the coupling of 
the top-quark to the $W$ boson and the bottom quark, $g_{tbW}$, as constants.\footnote{Therefore,
in contrast to the case of Higgs boson at the LHC, the assumed relation between 
onshell and offshell couplings is a much weaker one in the setup we consider here.}
Additionally, we only consider variations of the top-quark width up to $\pm 20$\% of the \sm{} value,
whereas in the case of the Higgs boson the experimental sensitivity corresponds to a width which is multiple 
times the \sm{} Higgs boson width.
Another difference is that here we consider ratios of cross sections in single and double resonant regions 
which receive contributions from phase-space where one or two reconstructed top quarks are `nearly' onshell.
In the Higgs boson case however,  contributions from onshell (resonant) Higgs boson production are compared 
with far-offshell Higgs boson production.
We discuss this in detail a little later. 

\subsection{Status and prospects of top-quark width measurements} \label{subsec:topquarkwidth}

Before proceeding to a detailed description of our proposed method, we first comment on the theoretical knowledge 
of $\gamt$ and its experimental measurement. 
At LO in the \sm{} the top-quark width is dominated by the decay into the $W$ boson and a $b$ quark,
which depends on the $g_{tbW}$-coupling as
\begin{align}
\Gamma(t\rightarrow Wb) = \left(\frac{g_{tbW}}{g}\right)^2
\frac{G_{\mu}\mt^3}{8\sqrt{2}\pi}\left(1-\frac{m_W^2}{\mt^2}\right)^2\left(1+\frac{2m_W^2}{\mt^2}\right)\,,
\label{eq:widthprop}
\end{align}
where $g$ denotes the electroweak coupling of SU(2)$_L$,
$G_{\mu}$ Fermi's constant and $m_t$ and $m_W$ the top-quark mass
and $W$ boson mass respectively (for simplicity we have set the bottom-quark mass 
to zero above).
Apart from NLO QCD corrections~\cite{Jezabek:1988iv,Czarnecki:1990kv,Li:1990qf} 
higher order QCD as well as electroweak corrections to $\Gamma(t\rightarrow Wb)$ are
known~\cite{Denner:1990ns,Eilam:1991iz,Czarnecki:1998qc,Chetyrkin:1999ju,Gao:2012ja,Brucherseifer:2013iv}. 
In the \sm{} $g_{tbW}$ can be written as $gV_{tb}$ where $V_{tb}$ is the corresponding CKM matrix element.
Since the branching fraction to a $W$ boson and a $b$ quark is almost $100$\%, the total 
top-quark width is almost linearly dependent on $g_{tbW}^2$.

Now we give a short summary of current measurements of $\gamt$.
Its value can be deduced from the measurement of the branching ratio BR$(t\rightarrow Wb)$
together with the partial decay width $\Gamma(t\rightarrow Wb)$.
The former can be accessed through the ratio $R={\rm BR}(t\rightarrow Wb)/\sum_{q=d,s,b}{\rm BR}(t\rightarrow Wq)$
measurable from top-pair production, which, being experimentally compatible with $R=1$~\cite{Agashe:2014kda},
points towards BR$(t\rightarrow Wb)\sim 100$\%. 
This measurement also implies strong bounds on non-SM top-quark decays such as $t\rightarrow H^+b$~\cite{Agashe:2014kda}.
The partial decay width can be indirectly determined through \eq{widthprop} by a measurement of $g_{tbW}$, 
on which we subsequently focus.
Whereas $g$ is known with great precision~\cite{Agashe:2014kda},
the CKM matrix element can either be deduced
from a global fit $V_{tb}=0.99914\pm 0.00005$~\cite{Agashe:2014kda}
assuming unitarity of the CKM matrix or from single top quark production 
(whereas top-quark pair production is insensitive to $g_{tbW}$).
The average of the single top quark production cross section
obtained by the Tevatron and the LHC experiments
leads to $|V_{tb}|=1.021\pm 0.032$~\cite{Agashe:2014kda},
which can be used to indirectly extract the top-quark width to an accuracy of order $100$\,MeV.
The coupling $g_{tbW}$ can be altered in models beyond the \sm{}
-- we direct the interested reader to \mycites{Batra:2006iq,Bernreuther:2008us}
for concrete examples in the context of a 2-Higgs-Doublet Model, the Minimal Supersymmetric Standard Model
or top-color assisted Technicolor. Their effects are most
dominant in the left-handed part of the coupling $g_{tbW}$
and are in the range of a few percent.
We point out that though $g_{tbW}$ may differ from its \sm{} value, this does not affect the validity 
of the assumption of equal on- and offshell couplings discussed in \subsec{gamt-offshell}.
For completeness we note that CDF also obtained direct bounds on $\gamt$, 
specifically, $1.10\,\text{GeV}<\gamt<4.05$\,GeV at $68$\% confidence level~\cite{Aaltonen:2013kna},
from template fits of the reconstructed top masses in $t\bar{t}$ events.

At a linear collider $\gamt$ can be directly deduced from top-quark pair production at threshold (see \cite{Moortgat-Pick:2015yla}
and references therein).
The dependence of the cross section on $\gamt$ is nicely illustrated in \mycite{Beneke:2015kwa}.
Furthermore, the forward-backward asymmetry in $e^+e^-\rightarrow t\bar t$ near threshold
shows a clear dependence on $\gamt$~\cite{Juste:2006sv}.
Refs.~\cite{Martinez:2002st,Horiguchi:2013wra,Seidel:2013sqa} report a projected accuracy of $20-30$\,MeV on $\gamt$ from top-quark 
pair production measurements at threshold.
In the continuum, due to the fine-resolution detectors and the cleaner environment at a linear collider, 
performing fits of the invariant-mass lineshape (reconstructed via the decay products of the top quark) provides a realistic 
method to precisely determine the top-quark width.
We will comment on this extraction method in \subsec{pert-uncert}, however we highlight that 
\mycites{Abe:2010aa,Seidel:2013sqa} estimate that, using reconstruction of the invariant mass at a linear collider, 
$\gamt$ can be determined with a precision of $60-220$\,MeV for $\sqrt{s}=500$\,GeV and an 
integrated luminosity of $100$\,fb$^{-1}$.

As we will describe in the following sections, taking carefully chosen ratios of measurements of offshell and 
onshell regions (these will be quantified below) can also provide access to the top-quark width.
The ratios are independent of explicit powers of the coupling $g_{tbW}$, in principle allowing one to
disentangle the coupling $g_{tbW}$ and the width $\gamt$.

%%%%%%%%%%%%%%%%%%%%%%%%%%%%%%%%%%%%%%%%%%%%%%%%%%%%%%%%%%%%%%%%%%%%%%%%%%%%%%%%%%%%%
\section{$e^+e^- \to W^+W^-b\bar{b}$ at NLO in QCD} \label{sec:proc-defn}
%%%%%%%%%%%%%%%%%%%%%%%%%%%%%%%%%%%%%%%%%%%%%%%%%%%%%%%%%%%%%%%%%%%%%%%%%%%%%%%%%%%%%

\begin{figure}
\centering
\begin{tabular}{c c}
\includegraphics[trim=4.2cm 20.2cm 8.4cm 4.0cm,clip,width=5.5cm]{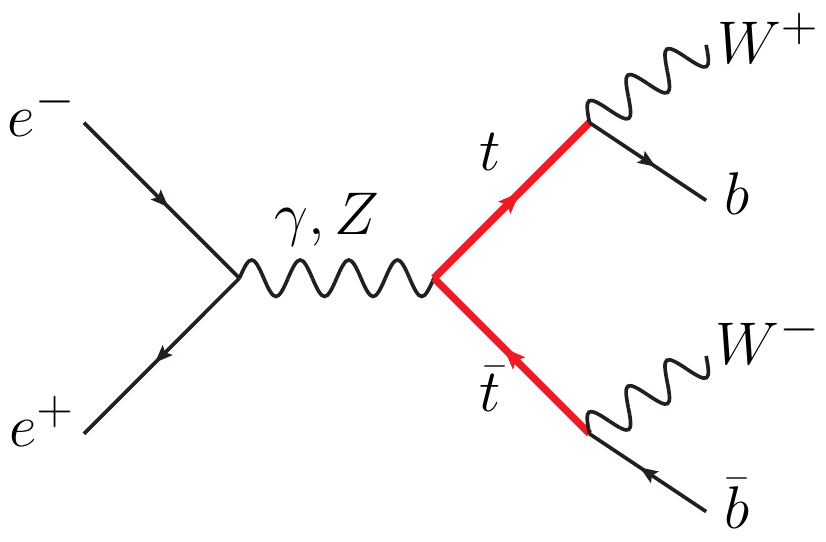} &
\raisebox{12pt}{ \includegraphics[trim=4.2cm 20.6cm 6.4cm 4.0cm,clip,width=6.9cm]{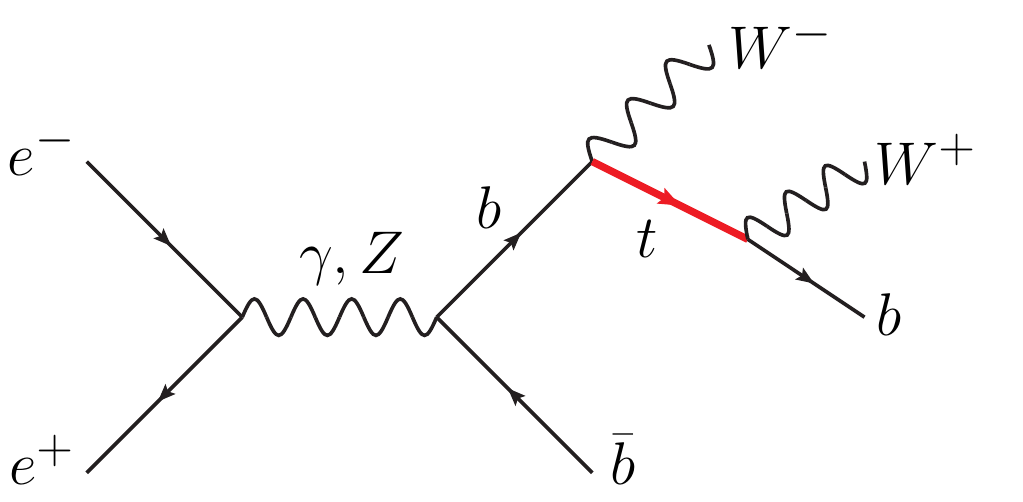} } \\[-12pt]
\hspace{-1.0cm} (a) & \hspace{-2.0cm} (b) \\[-12pt]
\includegraphics[trim=4.2cm 20.9cm 10.5cm 4.0cm,clip,width=5.9cm]{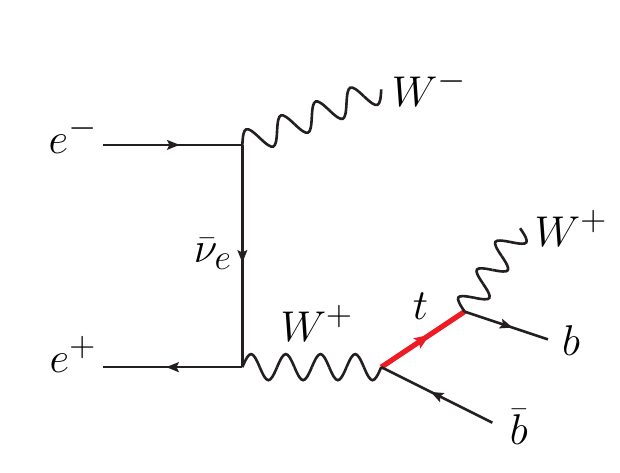} &
\raisebox{20pt}{\includegraphics[trim=4.2cm 20.9cm 6.4cm 4.0cm,clip,width=6.9cm]{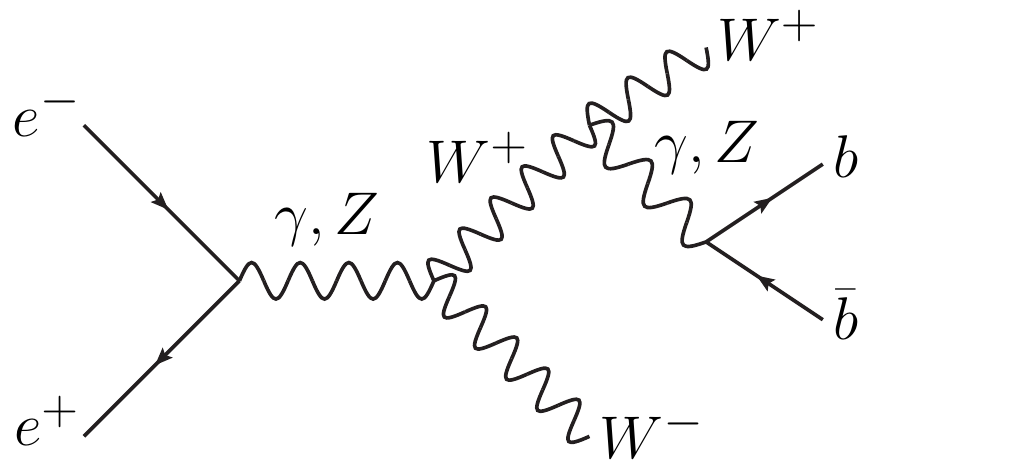}} \\[-12pt] 
\hspace{-1.0cm} (c) & \hspace{-2.0cm} (d) 
\end{tabular}
\vspace{-3mm}
\caption{Sample tree-level Feynman diagrams contributing to the full LO $e^+e^- \to W^+W^- b\bar{b}$ amplitude. 
These include (a) double resonant, (b) \& (c) single resonant and (d) non-resonant diagrams.}
\label{fig:lo-diagrams}
\end{figure}

The process we consider here is 
\begin{align} \label{eq:proc-def}
e^+e^- \to W^+ \, W^- \,  J_b \, J_{\bar{b}} + X
\,.\end{align}
Specifically, $J_b$ and $J_{\bar{b}}$ are bottom-flavoured jets containing at least a $b$ or a $\bar{b}$ parton respectively. 
The presence of top quarks is inferred by a reconstruction $b$-flavoured jets and $W$-bosons (the latter also determined via 
their decay products, leptons or jets, experimentally). 
The process is generated at fixed-order (LO and NLO-QCD) using the \mga 
code \cite{Alwall:2014hca} which uses {\sc MadLoop}~\cite{Hirschi:2011pa} for the evaluation of the one-loop matrix element
and {\sc MadFKS}~\cite{Frederix:2009yq} (based on FKS subtraction~\cite{Frixione:1995ms}) to handle the singular regions 
of the real corrections. 
Additionally, the complex-mass scheme \cite{Denner:1999gp,Denner:2005fg} is employed to consistently introduce the top-quark 
width. The bottom quark is considered to be stable and its mass is renormalized onshell.
We note that this process was first studied in \mycite{Lei:2008ii} and was briefly discussed in \mycite{Mattelaer:2014ewa}.
Recently the authors of the \whz{}~\cite{Moretti:2001zz,Kilian:2007gr} have also investigated this process at NLO-QCD 
in \mycite{Weiss:2015npa}. 
The Feynman diagrams contributing to this process include double, single and non resonant diagrams and at the amplitude-squared level all 
of these interfere with each other. Tree-level examples of these are shown in \fig{lo-diagrams}. 

The inclusive $W^+W^- b \bar{b}$ cross section is dominated by the double resonant $t\bar{t}$ contributions, namely, contributions
from diagrams such as diagram (a) in \fig{lo-diagrams} which contains two resonant top-quark propagators. This is particularly
the case near threshold. However, at centre of mass energies above threshold the relative contribution of single and non-resonant 
terms (arising from diagrams such as (b) \& (c) and (d) respectively of \fig{lo-diagrams}) increases \cite{Garcia:2014axa}. 
Single and non-resonant contributions also become very relevant below threshold
where the production of a resonant $t\bar{t}$ pair becomes kinematically suppressed, as discussed in \mycite{Batra:2006iq}.
Therefore, both in the continuum as well as below threshold, having a faithful description of the full $W^+W^-b\bar{b}$ final state is 
of great importance. 

As mentioned above, the complex-mass scheme is used to consistently introduce a complex mass at the Lagrangian level. 
This renormalization procedure replaces the bare top-quark mass, $m_{t,0}$ by a renormalized mass, $\mu_t$, and a 
counter-term, $\delta\mu_t$, both of which are complex,
\vspace{-3mm}
\begin{align}
m_{t,0} = \mu_t + \delta\mu_t\,,
\end{align}
where $\mu^2_t = m^2_t - i \mt \gamt$.
In this scheme the value of the top-quark width is considered as an input and the counter-term is chosen such that $\mu^2_t$ corresponds 
to the pole of the renormalized top-quark propagator. 
Formally, this means that if one uses fixed-order predictions with the complex-mass scheme to extract 
the top-quark mass, then the mass parameter one is sensitive to is the pole mass, defined as $\mt^2 = \text{Re}[ \mu_t^2 ]$.
Accordingly, the employed top-quark width is defined via $\mt\gamt = -\text{Im}[\mu_t^2]$.
The complex-mass scheme has already been used to compute NLO predictions for a number of processes involving unstable top quarks at hadron colliders
\cite{Bevilacqua:2010qb,Denner:2010jp,Denner:2012yc,Papanastasiou:2013dta,Frederix:2013gra,Cascioli:2013wga}. 

\subsection{Setup} \label{subsec:setup}
We summarize the parameter and analysis setup used throughout this paper in this subsection. 
The results we present, with the exception of the discussion in \subsec{com-energy}, are for a centre of mass energy of $\sqrt{s}=500$~GeV. 
We begin our investigation for unpolarised initial-state electrons and positrons, but extend our analyses to polarised
beams in  \subsec{polar}.
The numerical values of the relevant parameters used to produce our results are found in \tab{params}. 
We use \eq{widthprop}, including bottom-quark mass effects, to obtain our numerical
input value for the LO top-quark
width and the result of \mycite{Czarnecki:1990kv} to calculate
the NLO top-quark width.
In addition, we note that the assumption of a diagonal CKM matrix is made, namely we take $V_{tb}=1$.
The effect of a finite width of the $W$-boson is negligible for the observables we consider here, in particular since intermediate 
$W$-boson propagators are forced to be offshell by kinematics. At NLO QCD the cross section develops a dependence
on the renormalization scale $\muR$ (see \subsec{results}) and we employ a central scale choice of $\muR=\mt$ -- a standard scale
choice for the study of the $t\bar{t}$ process in $e^+e^-$ collisions. As we will see later, the inclusive cross section is 
very mildly dependent on this scale.\footnote{In the respective
hadron-collider process, $pp \to W^+W^-b\bar{b}$ choosing appropriate renormalization and factorization scales
is not as simple, particularly when trying to describe the single resonant contribution of the cross section. This is due to the different
underlying subprocesses governing the double and single resonant regions. For an in-depth discussion on this and on a possible scale-setting 
procedure for $pp$ collisions we refer the reader to \mycite{Cascioli:2013wga}.}

\vspace{0.2cm}
\begin{table}
\begin{tabular}{c c c c}
\hline
\multicolumn{4}{c}{Parameter Setup} \\
\hline 
$\mt = 173.2$~GeV & $m_b=4.75$~GeV & $m_W=80.385$~GeV & $m_Z=91.1886$~GeV \\
$m_H=125$~GeV & $\Gamma_Z=2.505$~GeV & $\gamh=4.21$~MeV & $G_{\mu}=1.1664\times10^{-5} \text{GeV}^{-2}$ \\ 
\hline 
\end{tabular}
\caption{Parameter choices.}
\label{tab:params}
\end{table}

Partons in the final state are clustered into (a maximum of three) jets using the $k_t$-algorithm, as implemented in \texttt{fastjet} \cite{Cacciari:2011ma}.
Tagging jets as $b$, $\bar{b}$ or light jets is done using the flavour information of partons in each jet that is available in our
parton-level analysis.
For most results we use a jet radius of $R_{\text{jet}}=0.5$. However, since the different combinations through which gluon radiation 
can be clustered play an important role in the structure of the invariant-mass distributions (see below), the jet-radius parameter, $R_{\text{jet}}$, 
is varied (enlarged) in order to better understand the extent to which this affects $\mt$ or $\gamt$ extractions. 
Minimal cuts of $p_T(J_b),p_T(J_{\bar{b}})  > 10$~GeV and $\lvert\eta(J_b)\rvert,\lvert\eta(J_{\bar{b}})\rvert < 4.5$ have been 
applied to the $b$-jets to define a typical fiducial region. 
This means that phase space points for which the $b$ and $\bar{b}$ partons are combined into 
the same jet are dropped in our analysis.

\subsection{Inclusive and differential results} \label{subsec:results}

\begin{figure}[h]
\begin{center}
\includegraphics[trim=0cm 0cm 0cm 1.3cm,clip,width=10.5cm]{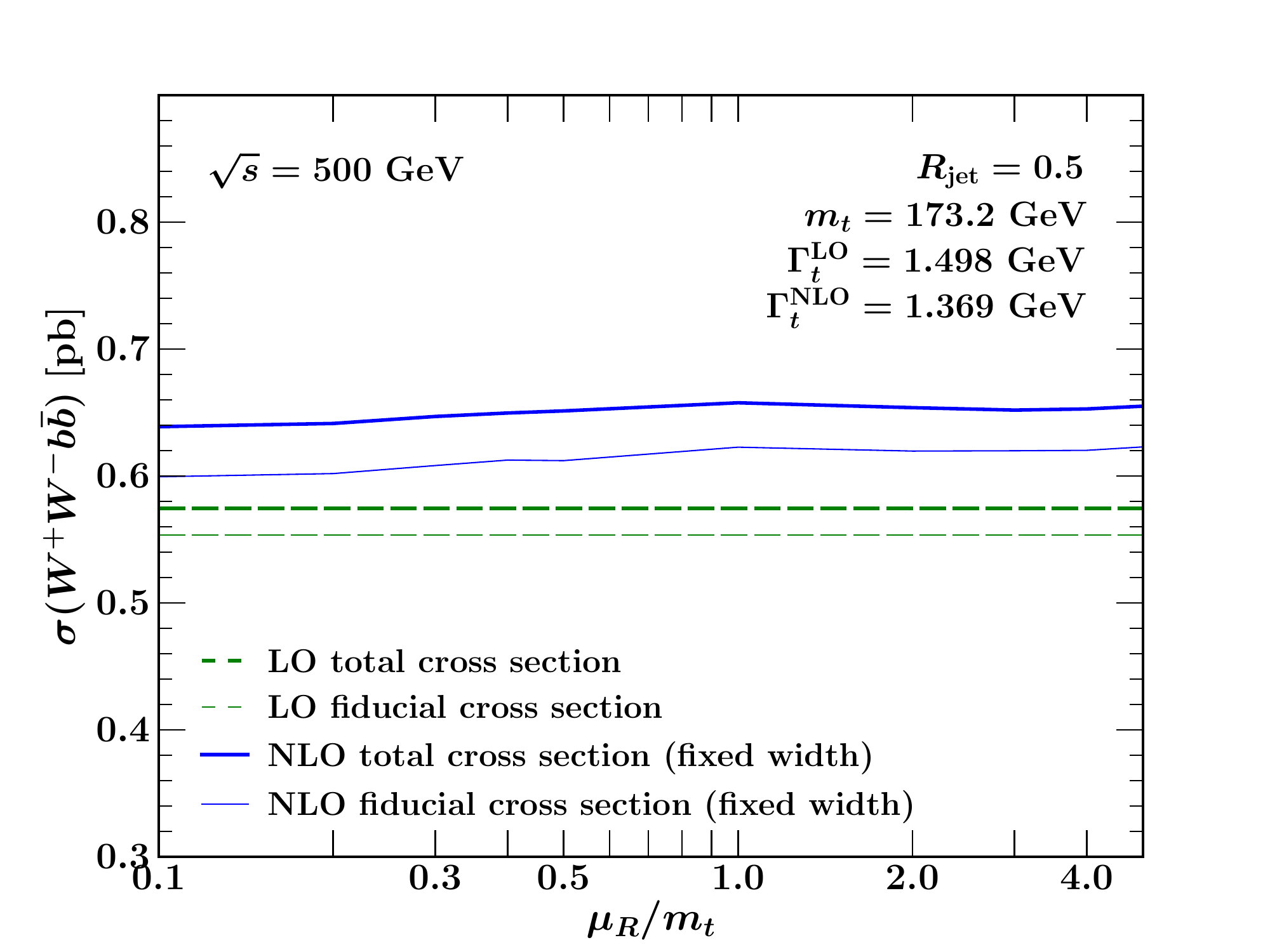}
\end{center}
\caption{Dependence on renormalization scale of inclusive (thick solid curves) and fiducial (thin solid curves) cross sections.
The dashed green curves indicate the LO cross sections whist the solid blue curves show the $\muR$-dependence of the NLO cross sections.}
\label{fig:xs-scaledep}
\end{figure}

We first briefly discuss the dependence the cross sections have on the renormalization scale $\muR$. We note that 
at LO, since the amplitudes do not depend on $\as$ there is no dependence on $\muR$. Of course, the NLO cross section
picks up a dependence on $\muR$ and in order to study this dependence we use a fixed renormalization scale $\muR=\xi \mt$ and 
vary $\xi \in [0.1,5]$. As illustrated in \fig{xs-scaledep}, both the fully-inclusive cross section as well as the 
fiducial cross section, defined according to the analysis cuts of \subsec{setup}, only have a mild dependence on $\muR$. 
We note however that the NLO corrections themselves are important, enhancing the LO numbers by around $10-12$\,\%. 

\begin{figure}[h]
\begin{tabular}{c c}
\hspace{-0.4cm}
\includegraphics[trim=0.2cm 0cm 1.85cm 0cm,clip,width=7.5cm]{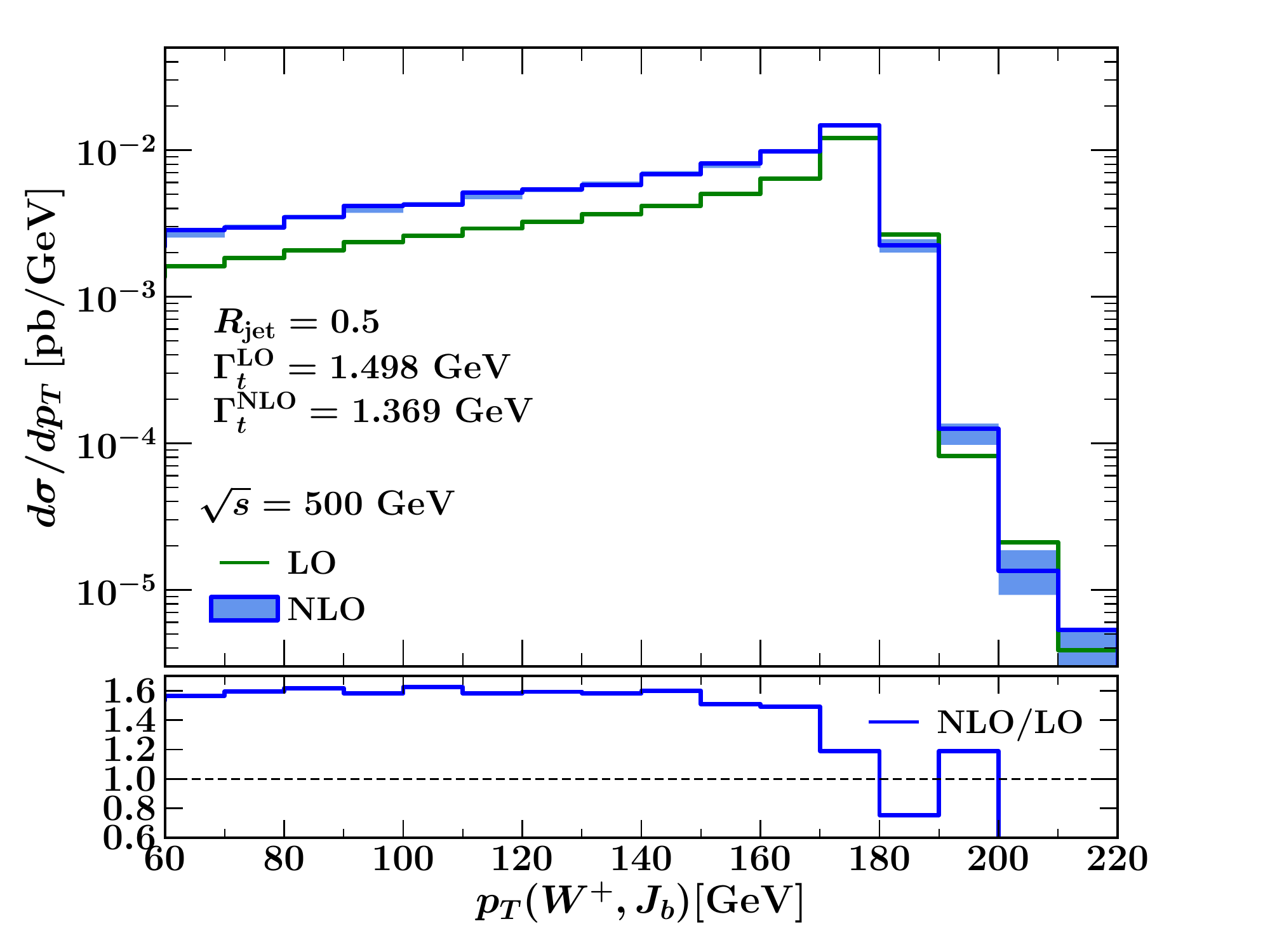} & 
\hspace{-0.4cm}
\includegraphics[trim=0.2cm 0cm 1.85cm 0cm,clip,width=7.5cm]{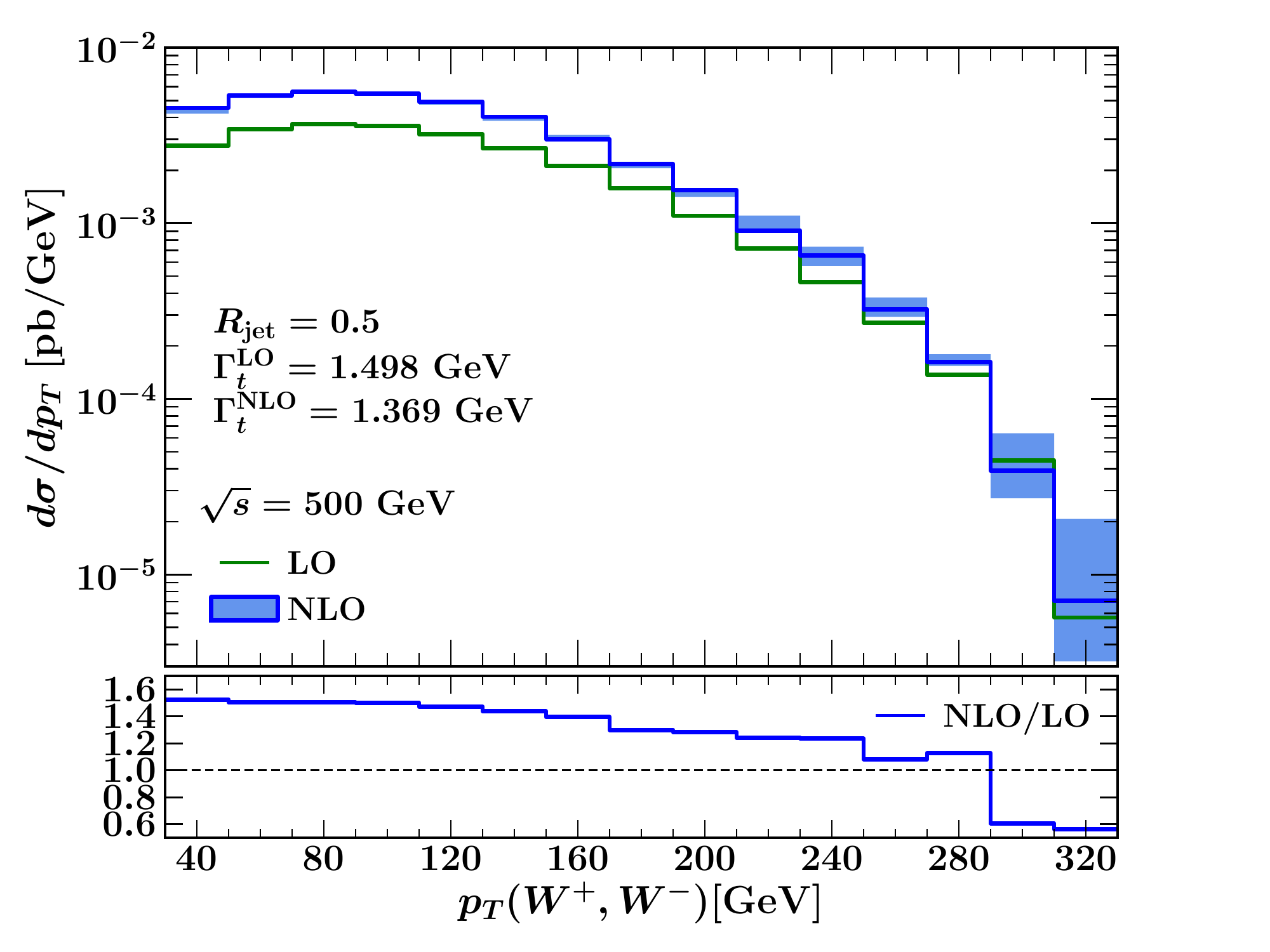} \\[-8pt] 
(a) & (b)
\end{tabular}
\caption{Distributions for the transverse momentum of (a) the reconstructed top quark, $p_T(W^+,J_b)$ and (b) the $W^+W^-$-pair, $p_T(W^+,W^-)$.
Upper panels: the green curves indicate the LO distributions whilst the blue band shows the NLO result, where the band is obtained by varying the renormalisation 
scale in the range $\muR \in [\mt/2,\, 2\mt]$.
Lower panels: the blue solid curves indicate the differential K-factor.}
\label{fig:pTwb-pTww}
\end{figure}

In \fig{pTwb-pTww} we show two example distributions, (a) the transverse momentum of the reconstructed top quark, $p_T(W^+,J_b)$ 
and (b) the transverse momentum of the $W^+W^-$ pair, $p_T(W^+,W^-)$. This is done to highlight that the code allows for 
the study of any infrared-safe differential observable that can be constructed using the full final state.
While it is not the purpose of this paper to discuss the effects of NLO corrections to the $e^+e^- \to W^+W^-b\bar{b}$
process, we indicate in the lower panels of \fig{pTwb-pTww} (a) and (b) that the NLO corrections do, for some observables,
lead to non-constant differential $K$-factors. Offshell and non resonant effects will also play an important role in
the tails of certain observables such as those in \fig{pTwb-pTww}. However, to quantify the role of such effects would require (at least) 
a comparison with the process in the narrow-width approximation, $e^+e^- \to t\bar{t}\to W^+W^-b\bar{b}$, at NLO (including NLO corrections 
to both production at decay subprocesses), which is beyond the scope of this paper. Such comparisons at NLO for hadron-collider
processes involving unstable top quarks can be found in 
\mycites{Falgari:2010sf,Falgari:2011qa,Denner:2010jp,Denner:2012yc,Falgari:2013gwa,Papanastasiou:2013dta,Cascioli:2013wga}. 

\subsection{Invariant mass of reconstructed top quarks} \label{subsec:distr-mwb}

We now examine the distribution for $M(W^+,J_b)$, explaining the structure behind the 
shapes of the curves at LO and NLO. A better understanding of this distribution will be key to 
explaining the patterns in the results we present in \subsec{pert-uncert} and \sec{offshell-width}.

\begin{figure}
\begin{tabular}{c c}
\hspace{-0.4cm}
\includegraphics[trim=0.2cm 0cm 1.85cm 0cm,clip,width=7.5cm]{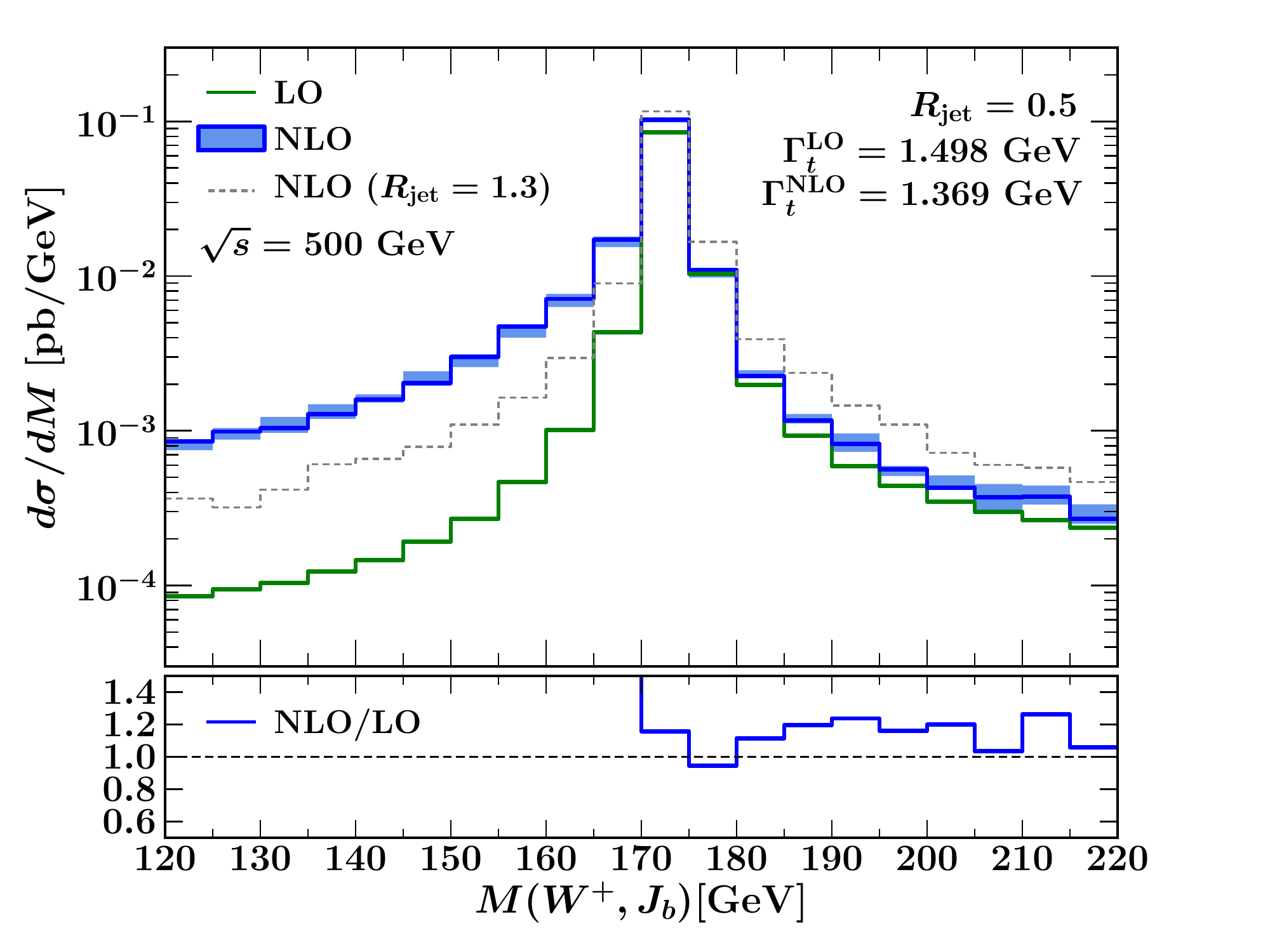} &
\hspace{-0.4cm}
\includegraphics[trim=0.2cm 0cm 1.85cm 0cm,clip,width=7.5cm]{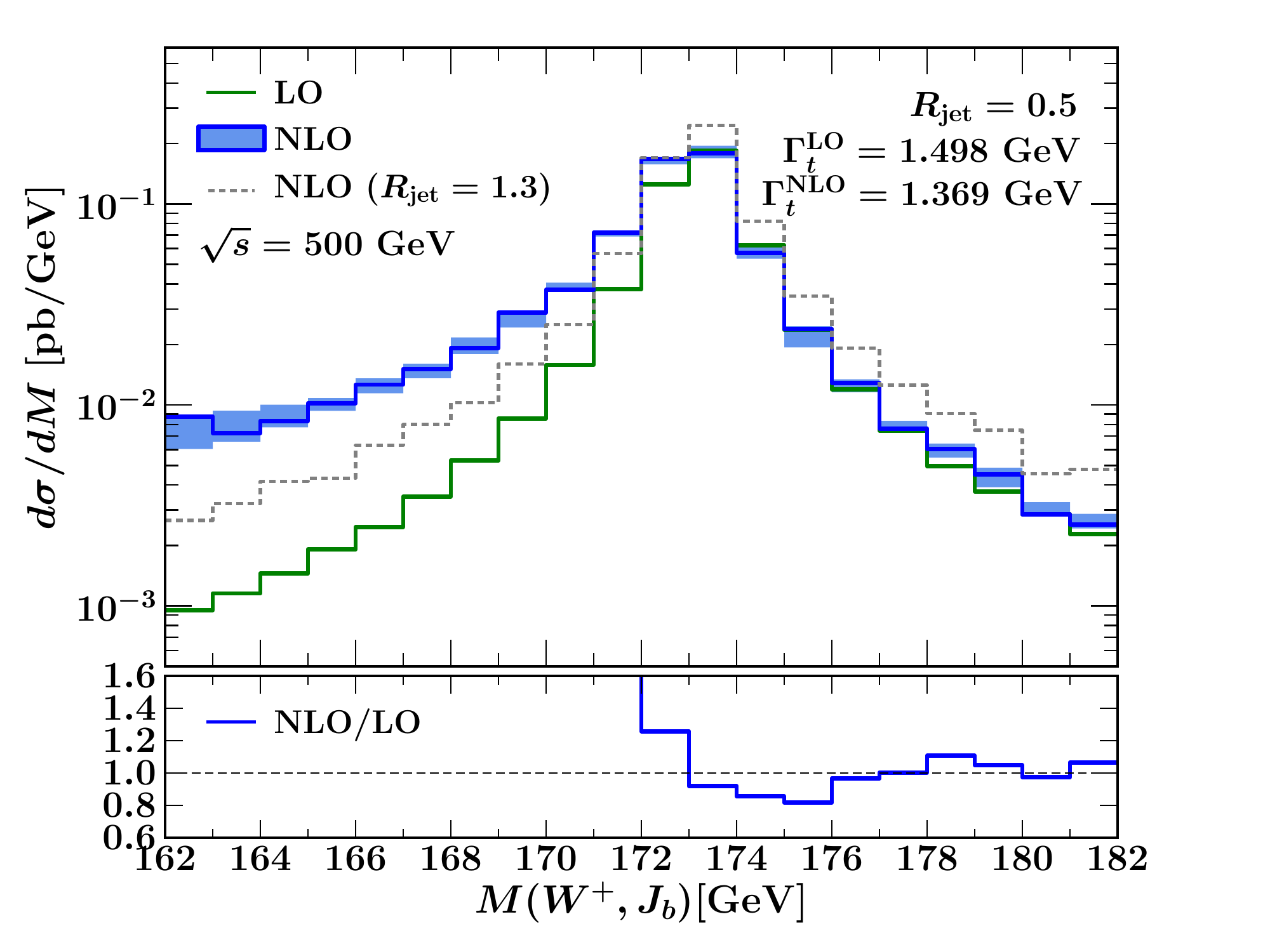} \\[-8pt] 
(a) & (b)
\end{tabular}
\caption{Distributions for the reconstructed top quark mass, $M(W^+,J_b)$, in the range (a) $M(W^+,J_b) \in [120,\,220]$~GeV and (b) $M(W^+,J_b) \in [162,\,182]$~GeV.
Upper panels: the green curves indicate the LO distributions whilst the blue band shows the NLO result, where the band is obtained by varying the renormalisation 
scale in the range $\muR \in [\mt/2,\, 2\mt]$. The green and blue solid curves show the results for a jet radius of $R_{\text{jet}}=0.5$ whilst the dashed gray 
curve shows the NLO distribution for $R_{\text{jet}}=1.3$.
Lower panels: the blue solid curves indicate the differential K-factor.}
\label{fig:mwb}
\end{figure}

In \fig{mwb} we plot the invariant-mass distribution of the reconstructed top quark, $M(W^+,J_b)$ at LO in green 
and at NLO in blue, for $\rjet=0.5$ in the ranges (a) $[120,220]$~GeV and (b) $[162,182]$~GeV. 
The shape of the LO curve around the peak, where the cross section is dominated by diagrams involving intermediate top quarks, 
is that of a standard Breit-Wigner distribution. Moving out towards the tails of the distributions, non resonant diagrams contribute 
to distort this shape. 
Going from LO to NLO one observes large differences between the LO and NLO results, with the LO curve lying outside the 
NLO scale uncertainty band, particularly for the region $M(W^+,J_b) < \mt$. 

\begin{figure}
\begin{center}
\includegraphics[trim=2.4cm 22.3cm 4.4cm 3.6cm,clip,width=14.5cm]{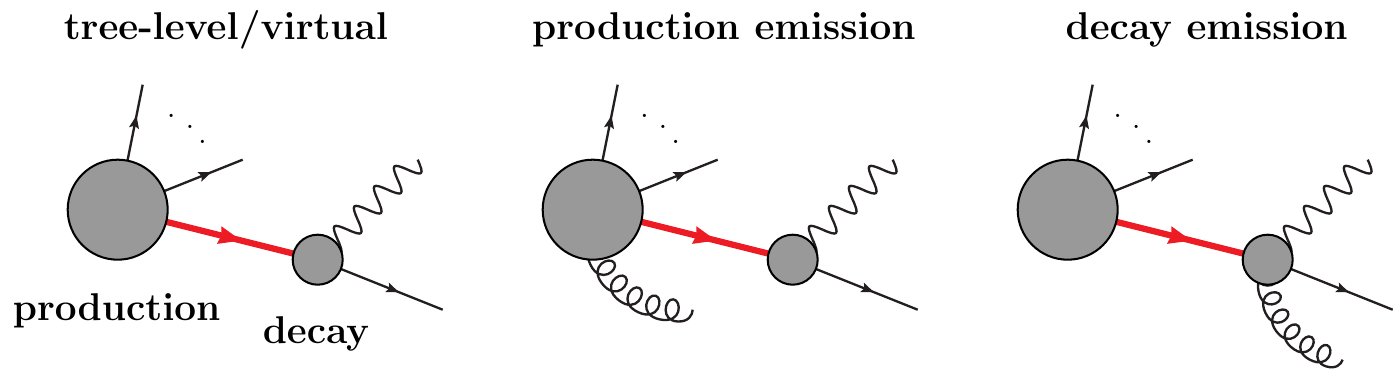} \\[-6pt]
(a) \hspace{4.0cm} (b) \hspace{4.0cm} (c)
\end{center}
\caption{Schematic diagrams indicating structure of amplitudes in limit of onshell top quark production and decay.}
\label{fig:res-str}
\end{figure}

To understand the reasons behind these large differences it is instructive to first consider the case of  
production and decay of a single onshell top quark. In this case at LO it is always the case that the intermediate top-quark 
momentum is $p_t=p_W+p_b$, as illustrated in \fig{res-str}(a), and all the cross section sits at $M(W^+,J_b)=\mt$. 
NLO virtual corrections to this process do not change the virtuality of the reconstructed invariant
mass, namely for virtual contributions we still have $M(W^+,J_b)=\mt$. 
In contrast, NLO real corrections can change the virtuality of the reconstructed top. In the onshell approximation
for the real-emission contributions we can either have that the intermediate top momentum is equal to $p_t=p_W+p_b$, 
for the case of an emission from the production subprocess, or that the top momentum is equal to $p_t=p_W+p_b+p_g$, in the case of an 
emission from the top-decay subprocess. These two cases are illustrated schematically in \fig{res-str} (b) and (c) 
respectively. 

When the momenta for the $b$ and $g$ momenta are fed to the jet algorithm the gluon momentum is either clustered
together with or separately from the $b$-parton. 
In the case where there is an emission from the production subprocess, then either the gluon is clustered with the 
$b$-parton, thus increasing the reconstructed mass $M(W^+,J_b) > \mt$, or the gluon is not clustered with the $b$-parton, 
and the invariant mass remains at $M(W^+,J_b)=\mt$.  
In the case of a gluon emission from the decay subprocess, then either the gluon is clustered together with the $b$-parton
meaning that the invariant mass remains at $M(W^+,J_b)=\mt$, or the gluon is clustered separately from the 
$b$-parton resulting in $M(W^+,J_b) < \mt$.
Since the contributions that change the virtuality of the reconstructed top
quark can only arise from real corrections at NLO (in the case of the narrow-width approximation that we consider here) 
they are positive contributions. In the case of onshell top production and decay this leads to tails forming away from 
the peak at $M(W^+,J_b)=\mt$.

Having understood the structure in the case of onshell top production, we move to the case of interest, namely offshell 
top-pair production, or $W^+W^-b\bar{b}$ production. Since intermediate top quarks are now generically offshell and additionally
non resonant terms contribute, the distribution for the invariant mass receives contributions both above and below $M(W^+,J_b)=\mt$  
starting at LO. The bulk of the cross section does still lie in or near the bin containing the point $M(W^+,J_b)=\mt$, indicating
that the resonant contributions are dominant. 
For this reason, the one-loop virtual corrections also do not change the structure of the LO curve significantly, even 
though there are many contributing corrections in addition to the one-loop corrections to the production and decay 
subprocesses that we considered in our toy-setup.

On the other hand the real corrections, as in the onshell case, can and do modify the LO shape of the reconstructed mass and 
the underlying reasons for this are precisely the same as those discussed in the case of an onshell top quark. 
Firstly, a positive contribution is expected for $M(W^+,J_b) < \mt$, due to emissions from an offshell resonant 
top, where $(p_W+p_b+p_g)^2 \sim \mt^2$, which are \emph{not} captured in the $b$-jet. 
Secondly a similar positive contribution is expected in the region $M(W^+,J_b) > \mt$, due to emissions that do not
change the virtuality of the intermediate resonant top, i.e. where $(p_W+p_b)^2 \sim \mt^2$, but which are 
however captured inside the $b$-jet. 

From the above discussion it is clear that precisely how the distribution for $M(W^+,J_b)$ is affected by the 
NLO corrections is dependent on how real radiation is clustered into jets and in particular the radius of the jets. 
In \fig{mwb} we also show the NLO distribution for $M(W^+,J_b)$ for $\rjet=1.3$ (typically used in many linear collider 
top-quark analyses), where we see that the effect of increasing the jet radius moves the NLO curve down for $M(W^+,J_b)<\mt$ and up for $M(W^+,J_b)>\mt$.
This is due to the fact that with a wider jet radius, on the one hand, one loses the gluon radiated from the top-decay subprocess outside
the jet (leading to the lowering of the tail $M(W^+,J_b)<\mt$) less often, but on the other hand, more frequently captures radiation from 
the top-production subprocess or elsewhere into the $b$-jet (resulting in the increase of the tail $M(W^+,J_b)>\mt$). 

Finally, we mention that given the impact of a single gluon emission on the shape of the distribution, it is evident that 
multiple gluon emissions during the parton-showering stage of a full event simulation will further affect the shape. 
In particular, parton-showering is expected to further broaden the lineshape of the reconstructed top quarks
and is certainly an effect worth additional investigation (though lies beyond the scope of this work).

\subsection{Uncertainty on $\mt$ and $\gamt$ extraction from kinematic reconstruction} \label{subsec:pert-uncert}

In this subsection we briefly comment on a method commonly used to extract both the top-quark mass and width in the continuum. 
This consists of a simple fit of a Breit-Wigner (BW) function to the reconstructed mass peak $M(W^+,J_b)$ or $M(W^-,J_{\bar{b}})$ 
(see for example \cite{Seidel:2013sqa}).
The method is usually applied to samples of simulated events where the underlying hard process is onshell top-quark pair 
production $e^+e^-\rightarrow t\bar t$.
These events are then supplemented with the corresponding LO decay of the top quarks by a parton shower and offshellness 
is inserted through a BW-smearing of the virtuality of the intermediate top-quarks. 
Since the exact shapes for the reconstructed masses can be predicted at both LO and NLO, and these are \emph{not} exact BW functions,
it is of interest to investigate to what extent fitting a BW function is a suitable method for extracting $\mt$ and $\gamt$ 
or whether significant errors are introduced in doing so. 
Here we focus on the extraction of the top-quark mass and width using the full process $e^+e^- \rightarrow W^+ W^- b\bar{b}$ at NLO in QCD.
Gluon emission at NLO in general broadens the peak, in particular at the invariant masses below the peak, and can thus potentially
pull the extracted mass towards a smaller central value. As discussed earlier, the shape of the reconstructed mass shows 
a strong dependence on the jet radius used in the definition of $b$-jets.

\begin{figure}[h]
\begin{center}
\includegraphics[width=10cm]{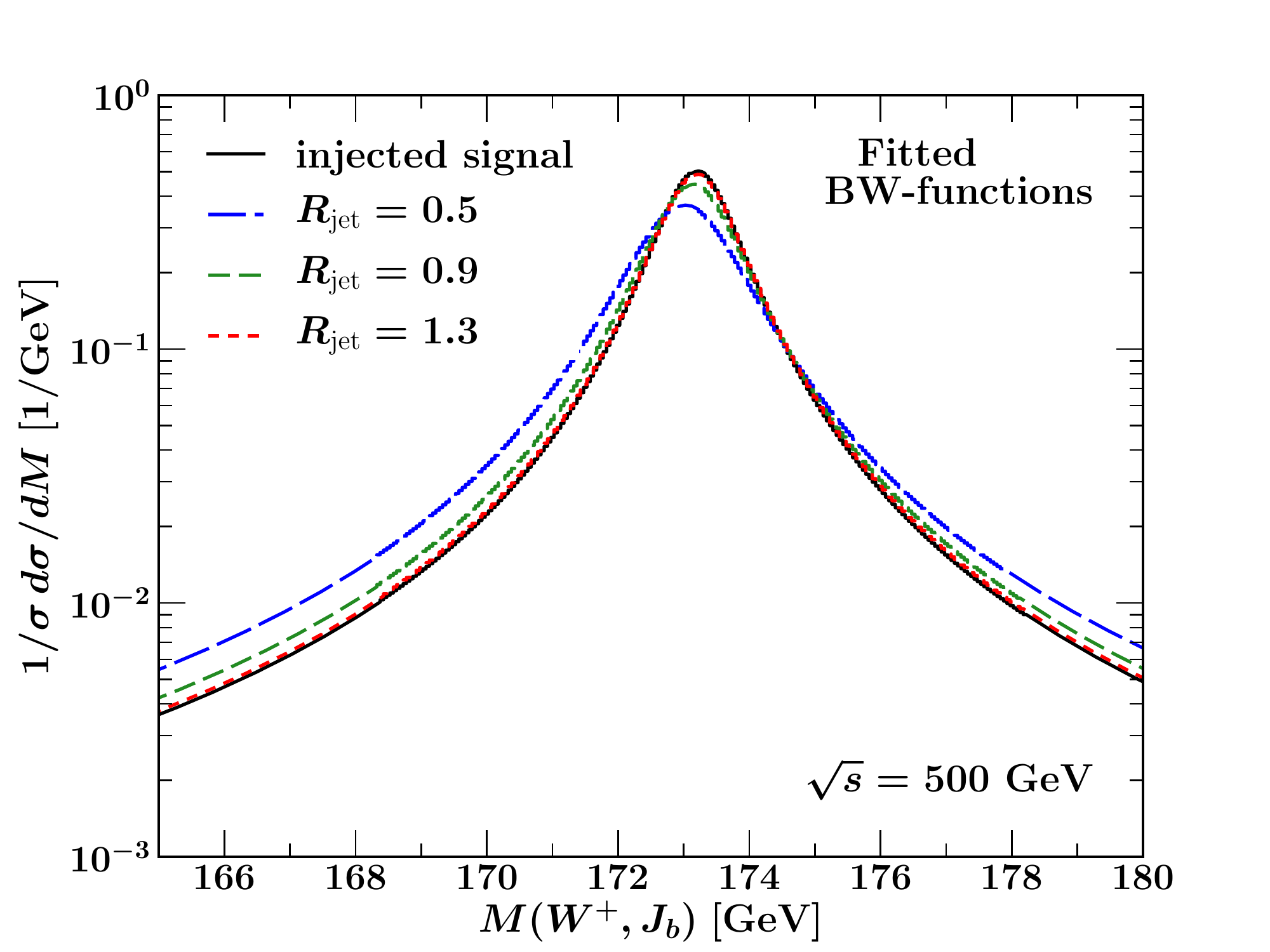}
\end{center}
\caption{Reconstructed Breit-Wigner shapes as a function of $M(W^+,J_b)$ for different employed values
of the jet radius. The curves are normalized to the total inclusive cross section. The injected signal
corresponds to the black, solid curve.}
\label{fig:mtgatrecons}
\end{figure}

We start with input values of $\mt = 173.2$~GeV and $\gamt=1.369$~GeV as input to \mga, using the latter to 
generate a distribution for $M(W^+,J_b)$ for $\rjet=\{0.5,0.9,1.3\}$. We then fit a BW function to 
these distributions extracting $\mt^{\text{meas}}$ and $\gamt^{\text{meas}}$ as those parameters for which 
the BW function best models the distribution. 
This is done using a least-squares method and the goodness-of-fit is comparable (and very good) in all 
three cases. Specifically, the standard deviation for the extracted values of $\mt^{\text{meas}}$ and $\gamt^{\text{meas}}$
is always below 60~MeV.
We note that we have performed this simple exercise assuming perfect $b/\bar{b}$-jet tagging and reconstruction of $W^+$ and $W^-$.
For $\rjet=0.5$ we extract a mass and width of $\mt^{\text{meas}}=173.00$~GeV and $\gamt^{\text{meas}}=1.92$~GeV, 
for $\rjet=0.9$ we find $\mt^{\text{meas}}=173.14$~GeV and $\gamt^{\text{meas}}=1.55$~GeV and  
for $\rjet=1.3$ we find $\mt^{\text{meas}}=173.20$~GeV and $\gamt^{\text{meas}}=1.30$~GeV. 
We see that with increasing $\rjet$ the extracted values for the mass and width approach the input values.
This is due to the fact that with increasing $\rjet$, gluon radiation from intermediate top decays is 
more likely to be clustered in a way that least distorts the LO BW shape near the resonance peak, as discussed
in \subsec{distr-mwb}. 
In \fig{mtgatrecons} we show the best-fit BW lineshapes as a function of the jet radius (in dashed blue, green and red) 
as well as the BW lineshape corresponding to the original input values used (in solid black).

To summarize, we would like to point out that a perturbative uncertainty of up to a few hundred MeV exists 
in the extraction of $\gamt$ using a fit of a BW function (which essentially models the LO invariant mass) 
to an NLO $M(W^+,J_b)$ distribution. The size of this depends on the jet radius and should 
be taken into account when performing such extractions. The origin of the uncertainty appears to 
be predominantly due to gluon emissions distorting the LO lineshapes.
This analysis is performed at fixed-order and despite the fact that parton showers capture some of the 
effects of hard radiation in the top-quark decay (and thus may decrease this uncertainty), we believe
that the systematic error on extracting $\gamt$ examined here will very likely remain until an extraction 
using the full NLO plus parton shower predictions of $e^+e^-\to W^+W^-b\bar{b}$ become available. 
Until these new tools are utilised it should be kept in mind that template or BW-fitting extractions 
of the width based on simulations using LO top-quark decays (such as those performed for example 
in \mycites{Abe:2010aa,Seidel:2013sqa}) ought to include a potentially important systematic error due 
to these missing higher-order effects.
We note that there is a corresponding, but smaller, uncertainty of about
200~MeV in the extraction of $\mt$.

We point out that it is known that the BW lineshape is additionally distorted by 
non-perturbative QCD effects. As explained in \mycites{Fleming:2007qr,Fleming:2007xt}, these 
can shift the extracted top-quark mass and width to larger values; an effect that increases
with the centre of mass energy. Control of such effects could be achieved through their encoding in 
universal soft functions.

%%%%%%%%%%%%%%%%%%%%%%%%%%%%%%%%%%%%%%%%%%%%%%%%%%%%%%%%%%%%%%%%%%%%%%%%%%%%%%%%%%%%%
\section{Sensitivity of offshell regions to the top-quark width} \label{sec:offshell-width}
%%%%%%%%%%%%%%%%%%%%%%%%%%%%%%%%%%%%%%%%%%%%%%%%%%%%%%%%%%%%%%%%%%%%%%%%%%%%%%%%%%%%%

In this section we investigate the extent to which the idea of using the cross section in offshell regions to probe or
place bounds on the total width can be applied to top-pair production when measurements on the full final state of $W^+W^-b\bar{b}$ are made. 
This possibility, which does not depend on fitting a particular functional form to a lineshape, is interesting to explore as an 
alternative handle on $\gamt$ in the continuum. Furthermore, the different choices one has in setting up this method in practice,
could be simultaneously exploited to consistently extract a precise value for $\gamt$.

Since there are two decaying top quarks, the method applied for the Higgs boson (discussed in \subsec{gamH-offshell}) has 
to be extended to consider the 
various resonance regions formed by both the top and antitop invariant masses. Given top quarks decay to $W$-bosons and 
$b$-quarks (measured as $b$-flavoured jets $J_b$ and $J_{\bar{b}}$ in experiments), the invariant 
masses one has to consider are those of \emph{reconstructed} top quarks, namely $M(W^+, J_b)$ and $M(W^-, J_{\bar{b}})$. 
We first try to provide some insight into the structure of the $W^+W^- b \bar{b}$ cross section by considering different 
resonance regions of the reconstructed masses, $M(W^+,J_b)$ and $M(W^-,J_{\bar{b}})$. 
The cross section can be divided up into double, single and non-resonant contributions, where these configurations
can be quantified according to the value of the measured invariant masses as follows: 
%%%
\begin{align} \label{eq:resonance-regions1}
\text{double resonant: } \;\;\; & M(W^+,J_b) \sim \mt^2 \;\; \text{and} \;\; M(W^-,J_{\bar{b}}) \sim \mt^2     \nonumber \\
\text{single resonant: } \;\;\; & M(W^+,J_b) \sim \mt^2 \;\; \text{and} \;\; \Big\{ M(W^-,J_{\bar{b}}) \ll \mt^2 \;\; \text{or} \;\; M(W^-,J_{\bar{b}}) \gg \mt^2 \Big\}  \nonumber \\
\text{single resonant: } \;\;\; & \Big\{ M(W^+,J_b) \ll \mt^2 \;\; \text{or} \;\; M(W^+,J_b) \gg \mt^2 \Big\}  \;\; \text{and} \;\; M(W^-,J_{\bar{b}}) \sim \mt^2          \nonumber \\
\text{non resonant: } \;\;\; & \Big\{ M(W^+,J_b) \ll \mt^2 \;\; \text{or} \;\; M(W^+,J_b) \gg \mt^2 \Big\}  \;\; \text{and} \;\; \nonumber \\
& \Big\{ M(W^-,J_{\bar{b}}) \ll \mt^2 \;\; \text{or} \;\; M(W^-,J_{\bar{b}}) \gg \mt^2 \Big\}\,. 
\end{align}
%%%
We note that the way in which we have chosen to define the different resonance regions depends 
on being able to faithfully tag a $b$-jet and a $\bar{b}$-jet.\footnote{The precise way
in which one divides up the phase space is of course arbitrary (see discussion later).}
Although challenging, discriminating between bottom quark and antiquark jets does appear to be possible at a Linear Collider, 
see for example the discussion in \mycite{Devetak:2010na} which proposes a novel quark-charge reconstruction algorithm to 
allow for such a selection. 
Since we consider the process with onshell $W$-bosons, we have also made the assumption of perfectly reconstructed $W$-bosons. 
These are clearly theoretical idealisations, but they nevertheless allow us to explore effects and features that would be present 
in a setup that additionally includes detailed simulations of other experimental effects (combinatorics, detector effects, etc).

The full matrix element comprising of the complete set of diagrams has a non-trivial dependence on 
several couplings. However, the coupling structure of the amplitudes giving the dominant contributions 
in the different resonant regions can be simplified. In the double resonant region, as defined 
in \eq{resonance-regions1}, the leading contributions are given by the double-resonant diagrams, and in this region 
the matrix element squared can be written as 
%%%
\begin{align}
\Big| \mathcal{M}^{\text{DR}} \Big|^2 = \Big| \sum_{V \in \{ \gamma,Z \} } g_{eeV} g_{ttV} A^{\text{DR}}_V  \Big|^2 \frac{g_{tbW}^4}{(\mt \gamt)^2}
+ \text{ subleading terms}
\,,
\label{eq:doubleres}
\end{align} 
%%%
with $A^{\text{DR}}_V$ denoting the amplitude for $e^+e^- \overset{V}{\rightarrow} t\bar{t} \rightarrow W^+ W^- b \bar{b}$
(see \fig{lo-diagrams}(a)).\footnote{The couplings $g_{ijk}$ arise from the Feynman rules for the vertices involving
the particles $i$, $j$ and $k$. }
In the above equation we have factored out the dependence of the amplitudes on $g_{tbW}$ as well as the denominators of the 
top-quark propagators which lead to the explicit factors of $\gamt$.\footnote{Note we have used the standard expansion 
$\frac{1}{(p_X-m_X)^2+m_X^2\Gamma_X^2} \to \frac{\pi}{m_X\Gamma_X}\delta(p_X^2-m_X^2) + \mathcal{O}\Big(\frac{\Gamma_X}{m_X}\Big)$ which 
holds in the limit $p_X^2 \to m_X^2$ applied at the matrix element squared level.} 

In each of the single-resonant regions defined in \eq{resonance-regions1}, the leading contributions arise from a linear 
combination of double and single resonant amplitudes and the matrix element squared in these regions can be written schematically as 
%%%
\begin{align}
\Big| \mathcal{M}^{\text{SR}} \Big|^2 &= \Big| \sum_{V \in \{ \gamma,Z \} } g_{eeV} g_{ttV} A^{\text{DR}}_V  
                                +  \sum_{V \in \{ \gamma,Z \} }\sum_{j \in \{ b, W \} } g_{eeV} g_{jjV} A^{\text{SR}}_{Vj} 
                                      + g_{eeW}^2 A^{\text{SR}}_W  \;  \Big|^2 \frac{g_{tbW}^4}{\mt \gamt} \nonumber \\
& \hspace{8.5cm} + \text{ subleading terms}
\,.
\label{eq:singleres}
\end{align} 
%%%
$A^{\text{SR}}_{Vj}$ and $A^{\text{SR}}_W$ are amplitudes arising from diagrams with a single resonant top quark, such 
as those in \fig{lo-diagrams}(b) and (c) respectively. 
The structure above can be understood as comprising of the single resonant component of the double resonant amplitudes (see \fig{lo-diagrams}(a))
and contributions from the single resonant amplitudes themselves (see \fig{lo-diagrams}(b) and (c)). 
Given that only one reconstructed top is resonant, the matrix element scales as $\gamt^{-1}$, however,
interestingly, these contributions scale with the same power of $g_{tbW}$ as for the double resonant case.

We now consider the ratio of the single resonant and double resonant region, which, following previous arguments
is insensitive to explicit powers of the coupling $g_{tbW}$. 
However, this ratio is (up to subleading terms) linearly dependent on the width, 
%%%
\begin{align} \label{eq:ratio-scaling}
\frac{\sigma^{\text{single-resonant}}}{\sigma^{\text{double-resonant}}} \propto \gamt\,. 
\end{align}
%%%
This means that the ratio we consider above indeed is a probe of $\gamt$ and is largely independent of variations
in $g_{tbW}$. 
Of course, $\gamt$ is itself sensitive to departures of $g_{tbW}$ from the \sm{} value, but may also 
change without modifications to the $g_{tbW}$ coupling and the latter possibility is one we wish to allow for.
The statement of \eq{ratio-scaling} holds at leading order and as we will see requires some refining when taking into account
higher orders. It is also dependent on the assumption that all couplings involved in the process take their \sm{} values.
Should the couplings $g_{ttV}$ differ from their \sm{} values, the ratio of single- and double-resonant
contributions will in general be altered and the determination of the top-quark width becomes more involved.
We note that subsequently when varying the top-quark width $\gamt$ we hold the coupling $g_{tbW}$ fixed
at its \sm{} value. A priori, a common rescaling of $g_{tbW} \to \xi g_{tbW}$ and $\gamt \to \xi^2 \gamt$ leaves the double resonant squared 
matrix element invariant, whereas the single resonant region is rescaled.
Therefore, \eq{ratio-scaling} provides the relevant counterpart ratio in the process $e^+e^-\to W^+W^-b\bar{b}$ that corresponds
to the ratio taken in studies for the Higgs boson width. 

The sets of amplitudes we have isolated in \eqs{doubleres}{singleres} are not themselves gauge-invariant; 
gauge invariance is restored once all subleading terms are included, which of course is the case for the full amplitudes 
we actually work with. 
However, it is nevertheless true that the amplitudes 
written down in these equations give the leading contributions in the double and single resonant regions. 
We note that approaches such as the pole expansion \cite{Stuart:1991xk,Aeppli:1993rs} or an effective theory expansion 
\cite{Beneke:2003xh,Beneke:2007zg,Falgari:2013gwa} provide gauge-invariant methods to compute cross sections to higher order 
in the different resonance regions without having to consider the full final-state amplitude. In such expansions the dominant 
terms in double and single resonant regions indeed receive their contributions from \eqs{doubleres}{singleres}. 
This therefore argues strongly in favour of the scaling of the ratio in \eq{ratio-scaling} as well as for it being virtually 
independent of $g_{tbW}$.

\subsection{Dividing up the cross section} \label{subsec:xs-structure}

In the previous subsection we argued that the ratio of single resonant to double resonant cross sections may 
provide a handle on the top-quark width. Here we set up a feasibility study of such a measurement, discussing
in particular the care required in choosing the resonance regions. 

According to the discussion above and in particular \eq{resonance-regions1} the cross section for $e^+e^-\to W^+W^-b\bar{b}$ 
can be divided up into double, single and non-resonant regions in the double differential distribution $d^2\sigma/dM(W^+,J_b)dM(W^-,J_{\bar{b}})$. 
We define reconstructed top and antitop quarks to be resonant if $M(W^+,J_b) \in [ M_{\text{min}}, M_{\text{max}} ]$~GeV
and $M(W^-,J_{\bar{b}}) \in  [ M_{\text{min}}, M_{\text{max}} ]$~GeV respectively. 
The cross section resonance regions can then be categorized as
\vspace{-3mm}
\begin{align} \label{eq:xs-resonance-split}
\text{double resonant   (DR):  } \;\; & M(W^+,J_b) \in [ M_{\text{min}}, M_{\text{max}} ] \text{ and } M(W^-,J_{\bar{b}}) \in [ M_{\text{min}}, M_{\text{max}} ] \nonumber \\[8pt]
\text{single resonant 1 (SR1):  } \;\; &  M(W^+,J_b) \in [ M_{\text{min}}, M_{\text{max}} ] \text{ and } M(W^-,J_{\bar{b}}) > M_{\text{max}} \nonumber \\
\text{or } & M(W^+,J_b) > M_{\text{max}} \text{ and } M(W^-,J_{\bar{b}}) \in [ M_{\text{min}}, M_{\text{max}} ] \nonumber \\[8pt]
\text{single resonant 2 (SR2):  } \;\; &  M(W^+,J_b) \in [ M_{\text{min}}, M_{\text{max}} ] \text{ and } M(W^-,J_{\bar{b}}) < M_{\text{min}} \nonumber \\
\text{or } & M(W^+,J_b) < M_{\text{min}} \text{ and } M(W^-,J_{\bar{b}}) \in [ M_{\text{min}}, M_{\text{max}} ]\,.
\end{align}
We use the notation $\sigma^{\text{DR}}$, $\sigma^{\text{SR1}}$, $\sigma^{\text{SR2}}$ to denote the cross section of the $W^+W^-b\bar{b}$ process 
in these phase space regions.
We have chosen not to list the non resonant regions above. This is because for the setup we study, the cross sections for these are 
negligible compared to those in the DR, SR1 and SR2 regions and therefore we do not consider them useful in this context. 
The boundaries $\big( M_{\text{min}}, M_{\text{max}} \big)$ determine the size of the resonant region for each reconstructed top quark.
The exact values are of course arbitrary and we vary them in three sets
$\big( M_{\text{min}}, M_{\text{max}} \big) \in \big\{ (165,180), (160,185), (155,190) \big\}$ GeV.\footnote{We note
that $\big( M_{\text{min}}, M_{\text{max}} \big) = (165,180)$ GeV (roughly) represent the boundaries outside which the effects of 
$\gamt$ in the Breit-Wigner propagator are smaller than $1$\,\%, i.e. for $M > 160~\text{GeV or } M>185 \text{GeV}$, 
we have that $((p^2_t-\mt^2)^2+\mt^2\gamt^2)/(p^2_t-\mt^2)^2 < 1.01$. }

The reason for having two separate single resonant (SR1 and SR2) regions has to do with higher-order 
corrections to the cross section. 
As we explained in \subsec{distr-mwb} and is clearly visible in \fig{mwb}, in the region of 
$M(W^+,J_b), M(W^-,J_{\bar{b}})<M_{\text{min}}$ the cross section is highly sensitive to additional gluon radiation from the 
decay products of a resonant intermediate top quark.
This means that the SR2 region will tend to receive very large NLO corrections from double resonant real contributions in which 
gluon emissions are not captured in the appropriate $b$ or $\bar{b}$ jet. 
As such the SR2 region is likely to suffer from a significantly larger renormalization scale dependence than the SR1 region. 
It is therefore, in practice, much less sensitive to variations in the top-quark width than a LO analysis 
may naively find.
Since the size of this effect is dependent on the choice made for $R_{\text{jet}}$ we investigate the impact of varying the jet radius
has on the final results.  

We study the impact on the structure of the cross section when varying the \sm{} top-quark width by $\pm20\%$
and fixing the coupling $g_{tbW}$. 
The LO cross section is independent of $\muR$ since the tree-level diagrams do not depend on $\as$.
The NLO cross section does however depend on the renormalization scales and this dependence must be 
quantified if a reliable estimate of the sensitivity to $\gamt$ is to be made.

\begin{figure}
\begin{tabular}{c c}
\hspace{-1.0cm} \includegraphics[trim=0cm 0cm 0cm 1.3cm,clip,width=8.4cm]{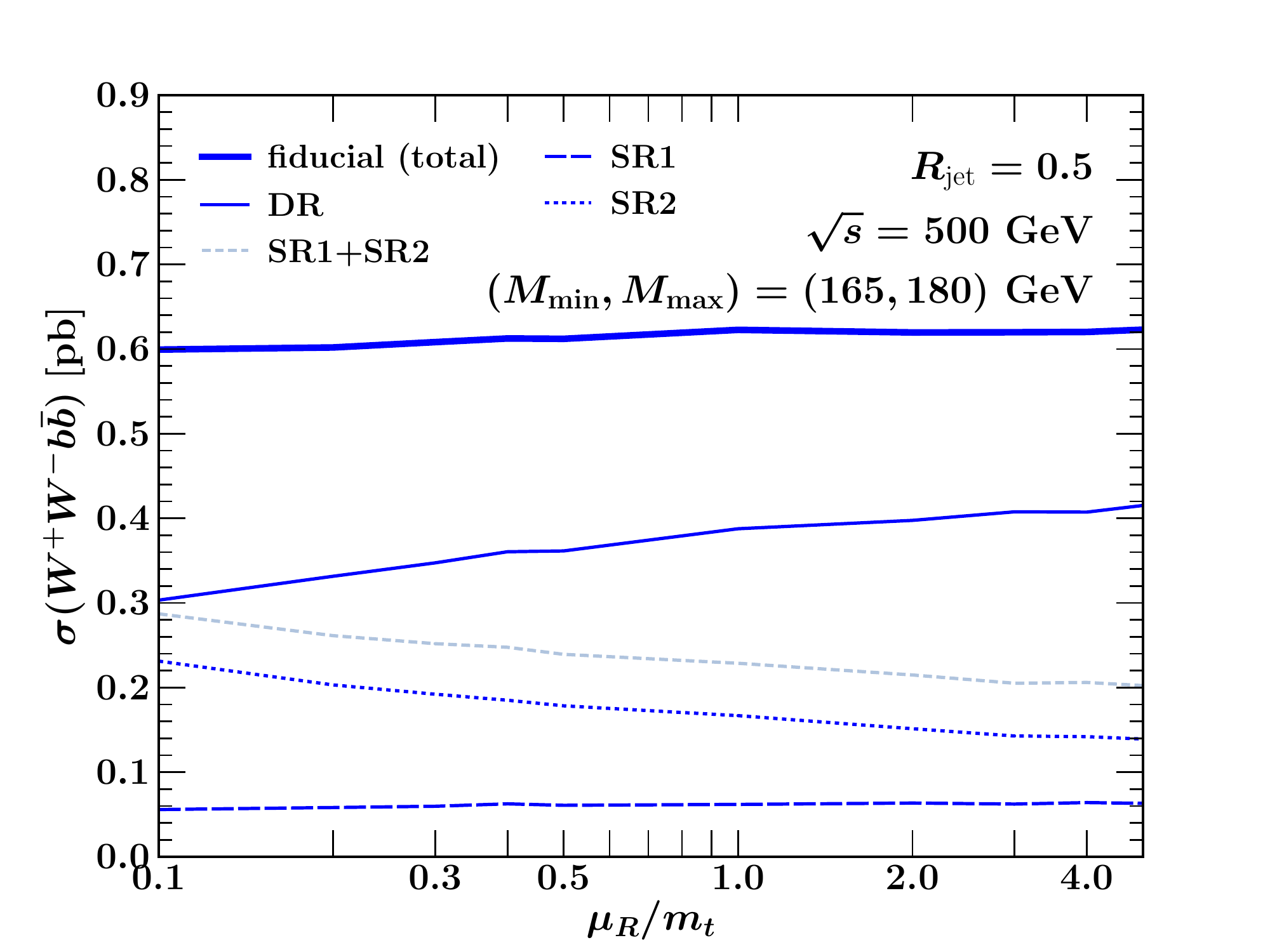}  
& \hspace{-1.2cm} \includegraphics[trim=0cm 0cm 0cm 1.3cm,clip,width=8.4cm]{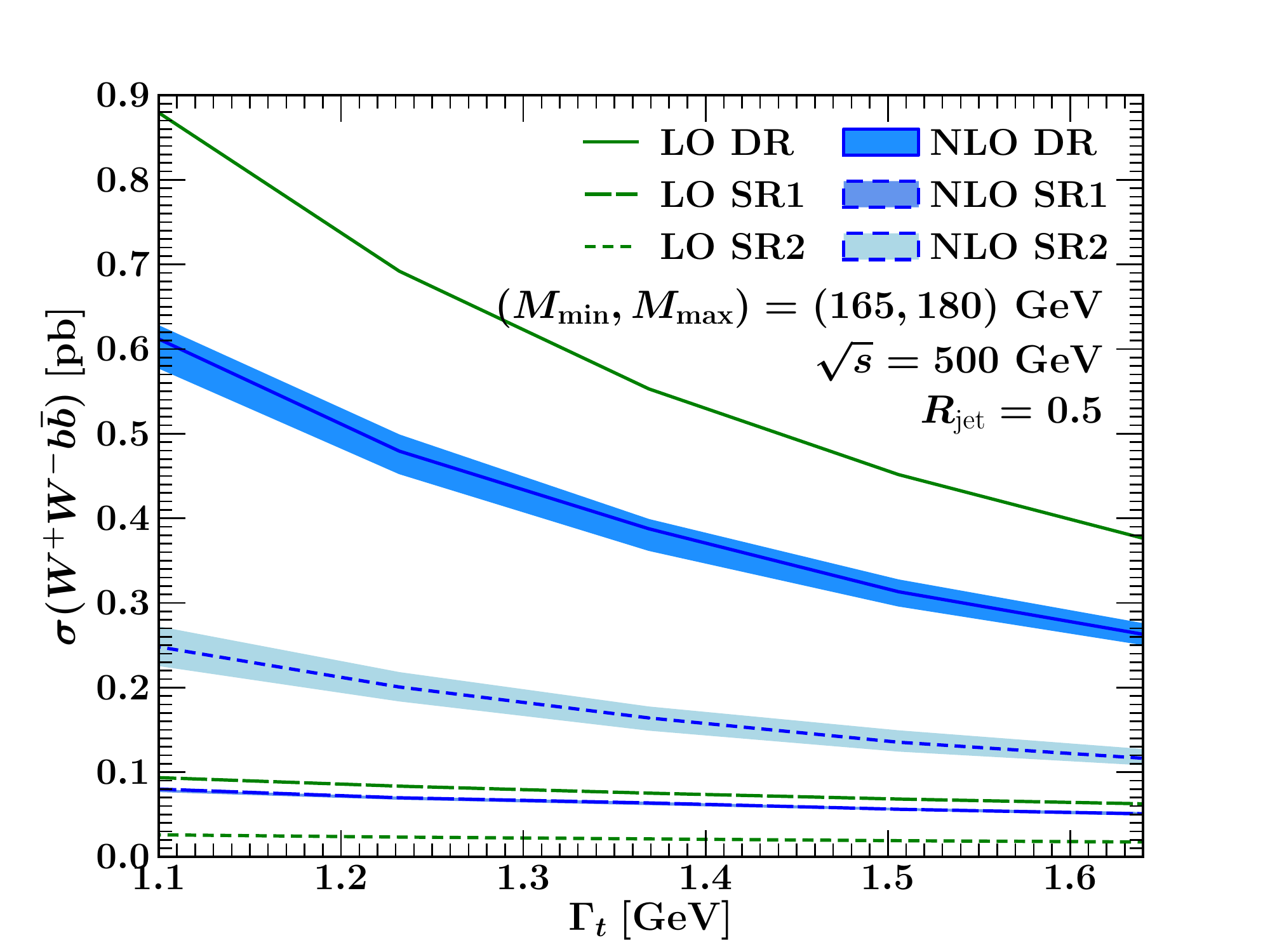} \\[-5pt]
(a) & (b)
\end{tabular}
\vspace{-3mm}
\caption{(a) Dependence on the renormalization scale of the NLO cross sections in the double and single resonant regions as defined in \eq{xs-resonance-split}.
(b) Dependence of LO and NLO DR, SR1 and SR2 cross sections on the top-quark width. The plots above illustrate the results for the choice 
$\rjet=0.5$ and $M_{\text{min}}=165$~GeV and $M_{\text{min}}=180$~GeV.}
\label{fig:xs-str-scaledep}
\end{figure}

In \fig{xs-str-scaledep}(a) we plot the dependence on $\muR$ of the fiducial cross section (thick solid blue), as well as the 
$\muR$-dependence of its double and resonant sub-regions: DR (thin solid blue), SR1 (long dashed blue) and SR2 (dotted blue). 
As also seen in \fig{xs-scaledep}, the total fiducial cross section is only very mildly dependent on $\muR$. 
However, \fig{xs-str-scaledep}(a) reveals that both DR and SR2 cross sections carry a dependence on $\muR$ and, moreover, that their
dependence goes in opposite directions. This crucially means that the ratio SR2/DR has a large dependence on $\muR$, thus making
it essentially insensitive within uncertainties to the relatively small variations in $\gamt$ we consider. 
On the other hand, the SR1 region is largely independent on $\muR$, making the ratio SR1/DR a potentially good one for probing $\gamt$. 
These patterns will be confirmed in the figures that follow. 
We note that similar conclusions as for the SR2 region hold for the total single resonant region SR1+SR2 since the latter
is dominated in size and in scale-dependence by the SR2 region (see the short dashed gray curve in \fig{xs-str-scaledep}(a)).	
For this reason we do not consider the full single resonant region any further. 

Figure \ref{fig:xs-str-scaledep}(b) shows the $\gamt$-dependence of the DR, SR1 and SR2 regions at LO (green) and NLO (blue). 
The clear $1/\gamt^2$ behaviour of the double resonant region both at LO and NLO illustrates that while varying the 
top-quark width we hold $g_{tbW}$ fixed. 
This also illustrates our earlier arguments that the NLO corrections to the SR2-region cross section are indeed very large, whilst 
the corrections to SR1 show a better behaviour.
Whilst the actual size of NLO corrections in each region is dependent on $\rjet$, 
these observations are generically true over all values of $\rjet$ that we have considered (i.e. $\rjet \in [0.5,1.3]$).
As also anticipated the $\sigma^{\text{SR2}}$ contribution shows a sizeable dependence on $\muR$, whilst $\sigma^{\text{SR1}}$ 
appears relatively unaffected by the variation. 

With an understanding of the structure of NLO corrections as well as the scale dependence of the various resonance regions 
we can now move to studying the ratios of interest. Specifically we examine the cross section ratios: 
%%%
\begin{align} \label{eq:xs-ratios}
\frac{\sigma^{\text{SR1}}}{\sigma^{\text{DR}}} \hspace{0.4cm} \text{ and }  \hspace{0.4cm}
\frac{\sigma^{\text{SR2}}}{\sigma^{\text{DR}}}
\,.\end{align} 
%%%
In order to verify the independence of our ratios on the coupling $g_{tbW}$, we have checked numerically that 
varying $g_{tbW}$ by $\pm10\%$ while keeping $\gamt$ fixed, indeed leaves the ratios unchanged.

\begin{figure}[H]
\begin{tabular}{c c}
\hspace{-1.0cm} \includegraphics[trim=0cm 0cm 0cm 1.3cm,clip,width=8.4cm]{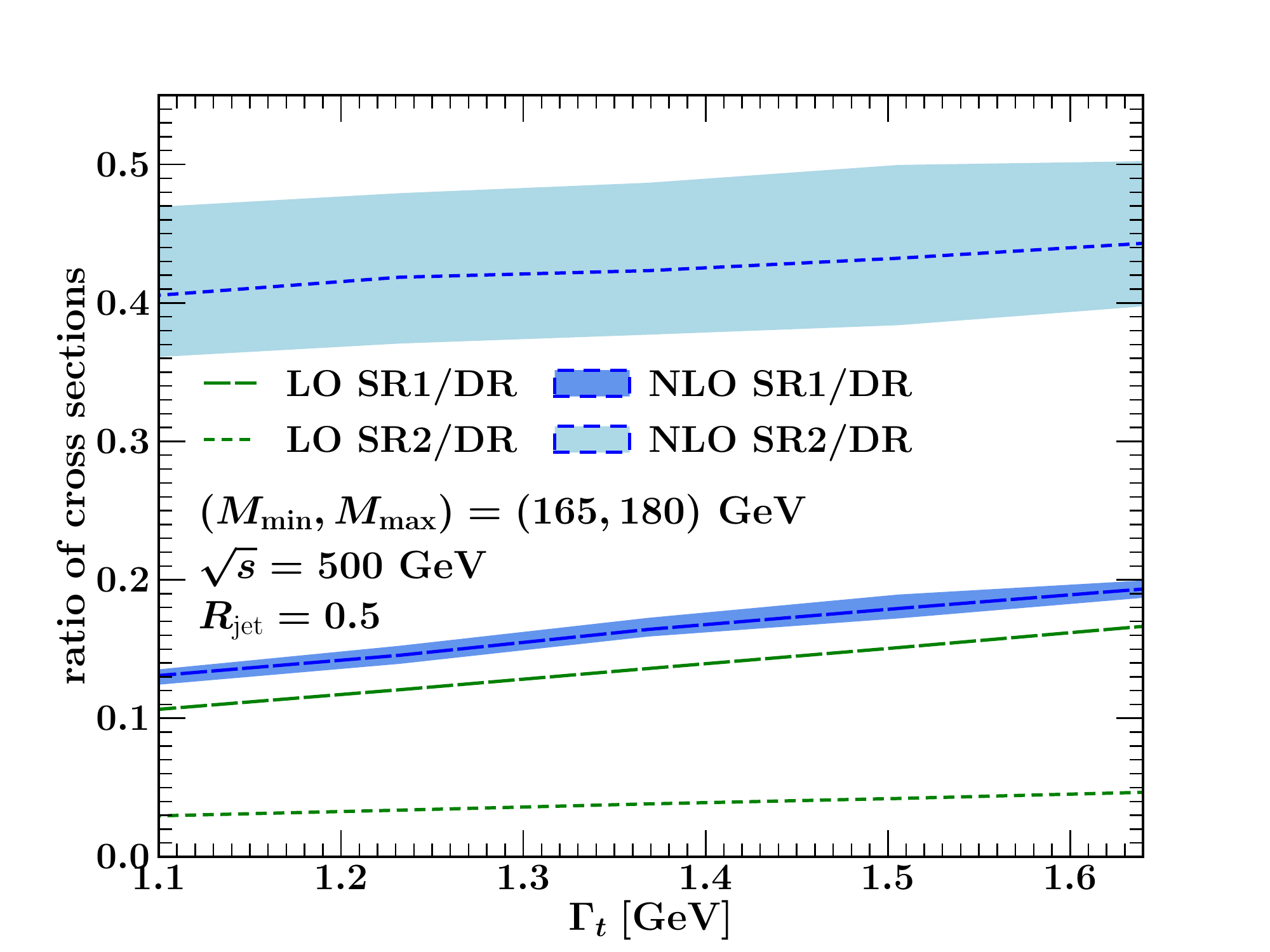}  
& \hspace{-1.2cm} \includegraphics[trim=0cm 0cm 0cm 1.3cm,clip,width=8.4cm]{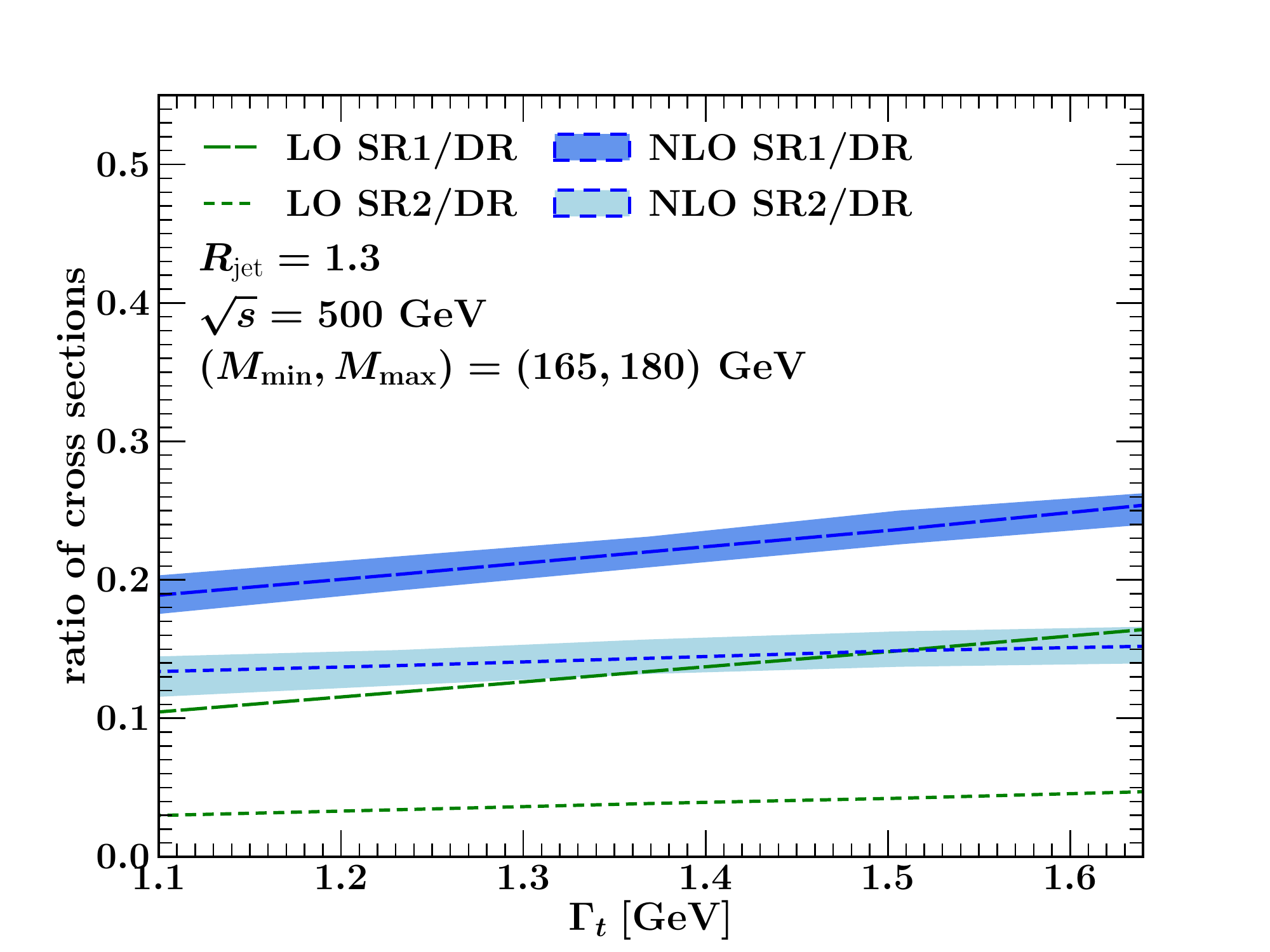} \\[-5pt]
(a) & (b)\\[5pt]
\hspace{-1.0cm} \includegraphics[trim=0cm 0cm 0cm 1.3cm,clip,width=8.4cm]{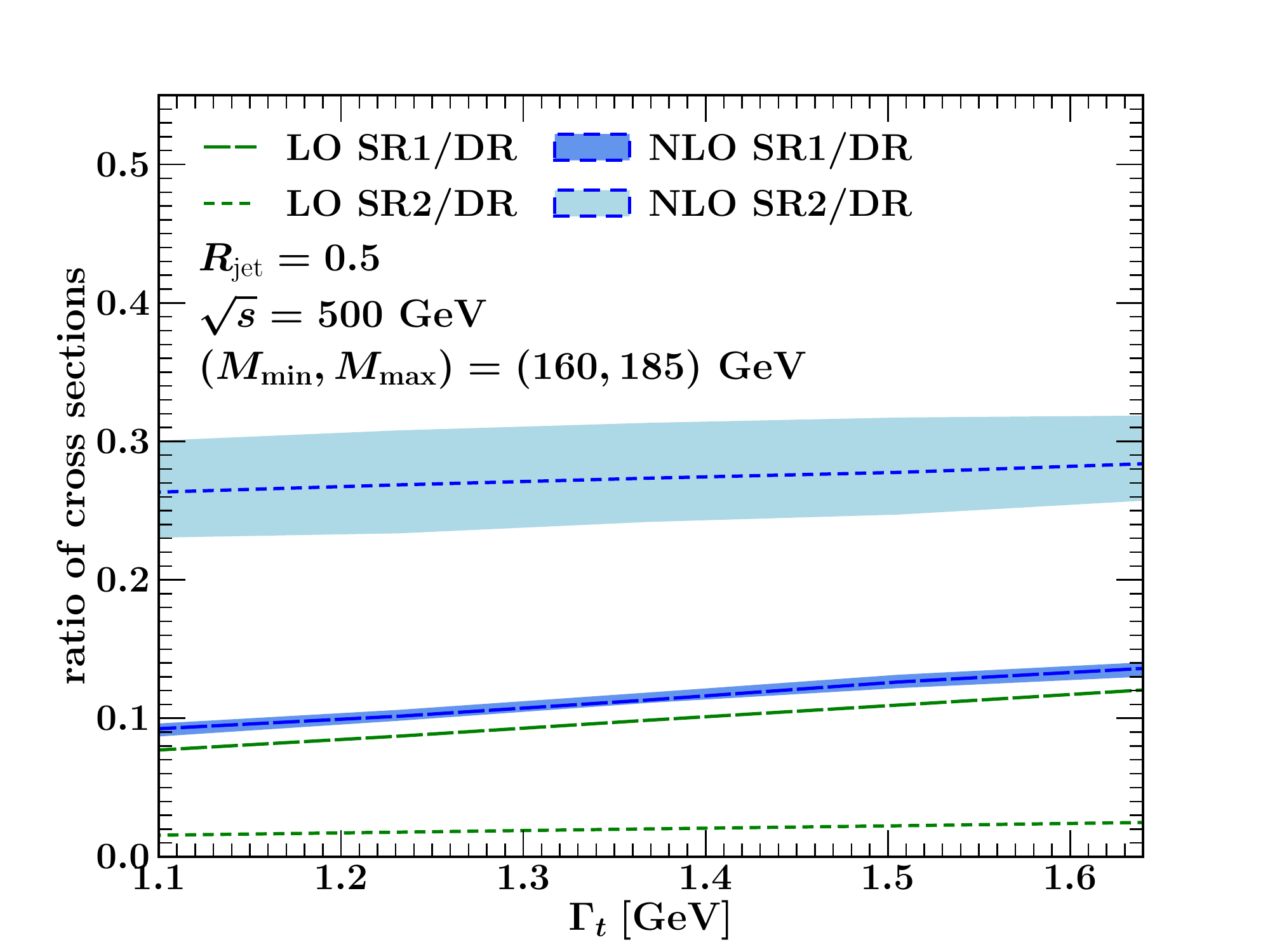}  
& \hspace{-1.2cm} \includegraphics[trim=0cm 0cm 0cm 1.3cm,clip,width=8.4cm]{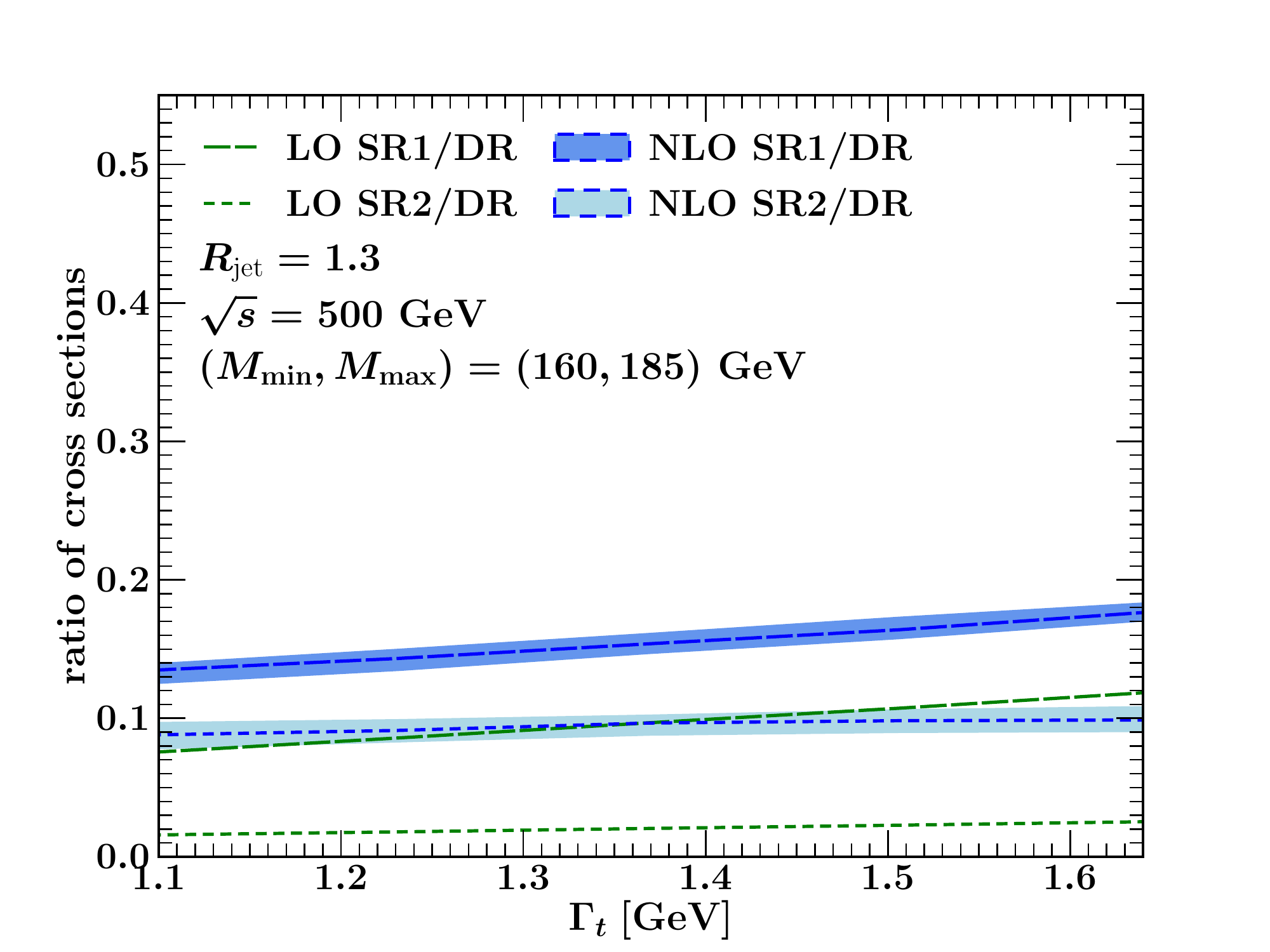} \\[-5pt]
(c) & (d)\\[5pt]
\hspace{-1.0cm} \includegraphics[trim=0cm 0cm 0cm 1.3cm,clip,width=8.4cm]{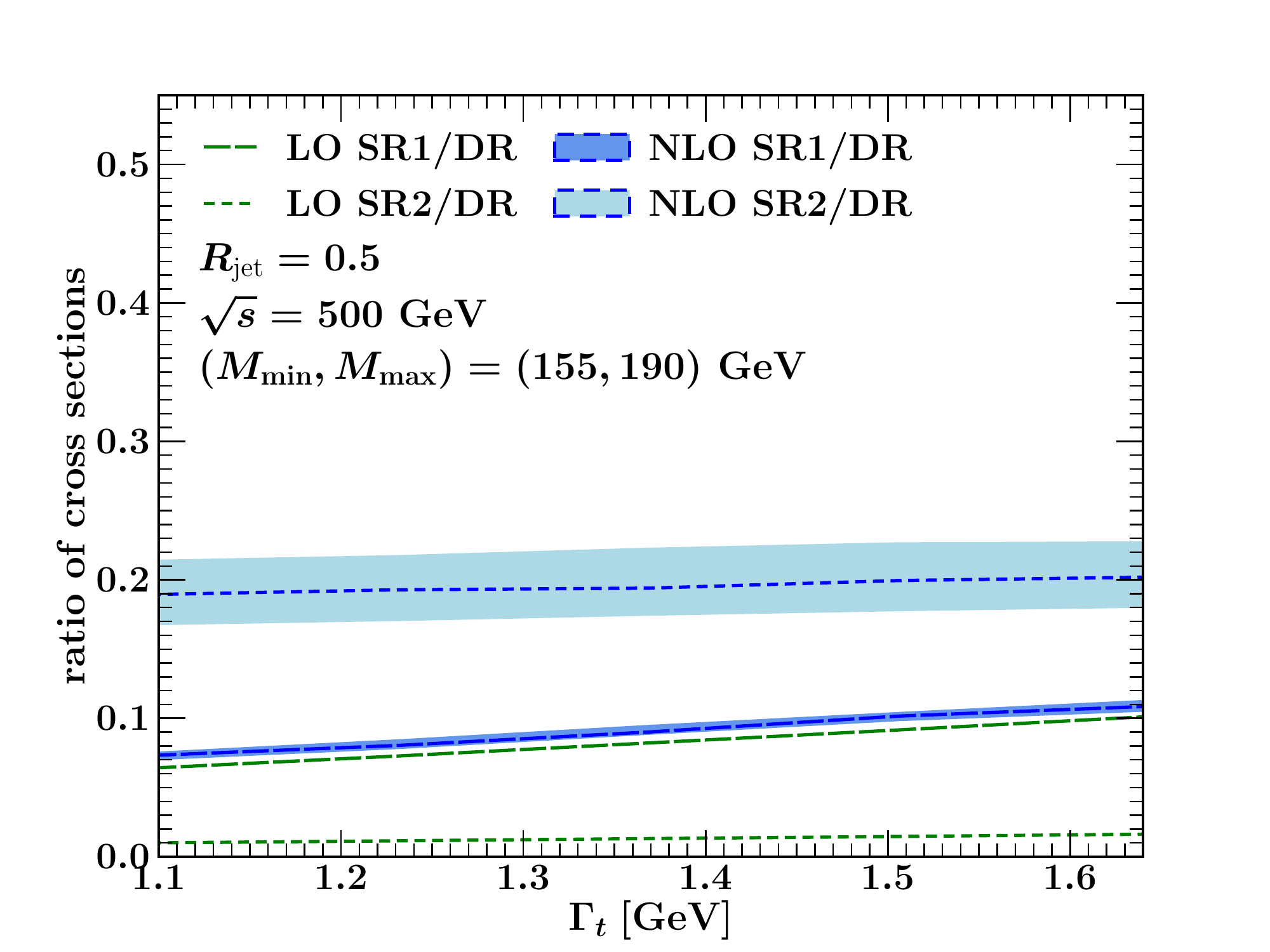}  
& \hspace{-1.2cm} \includegraphics[trim=0cm 0cm 0cm 1.3cm,clip,width=8.4cm]{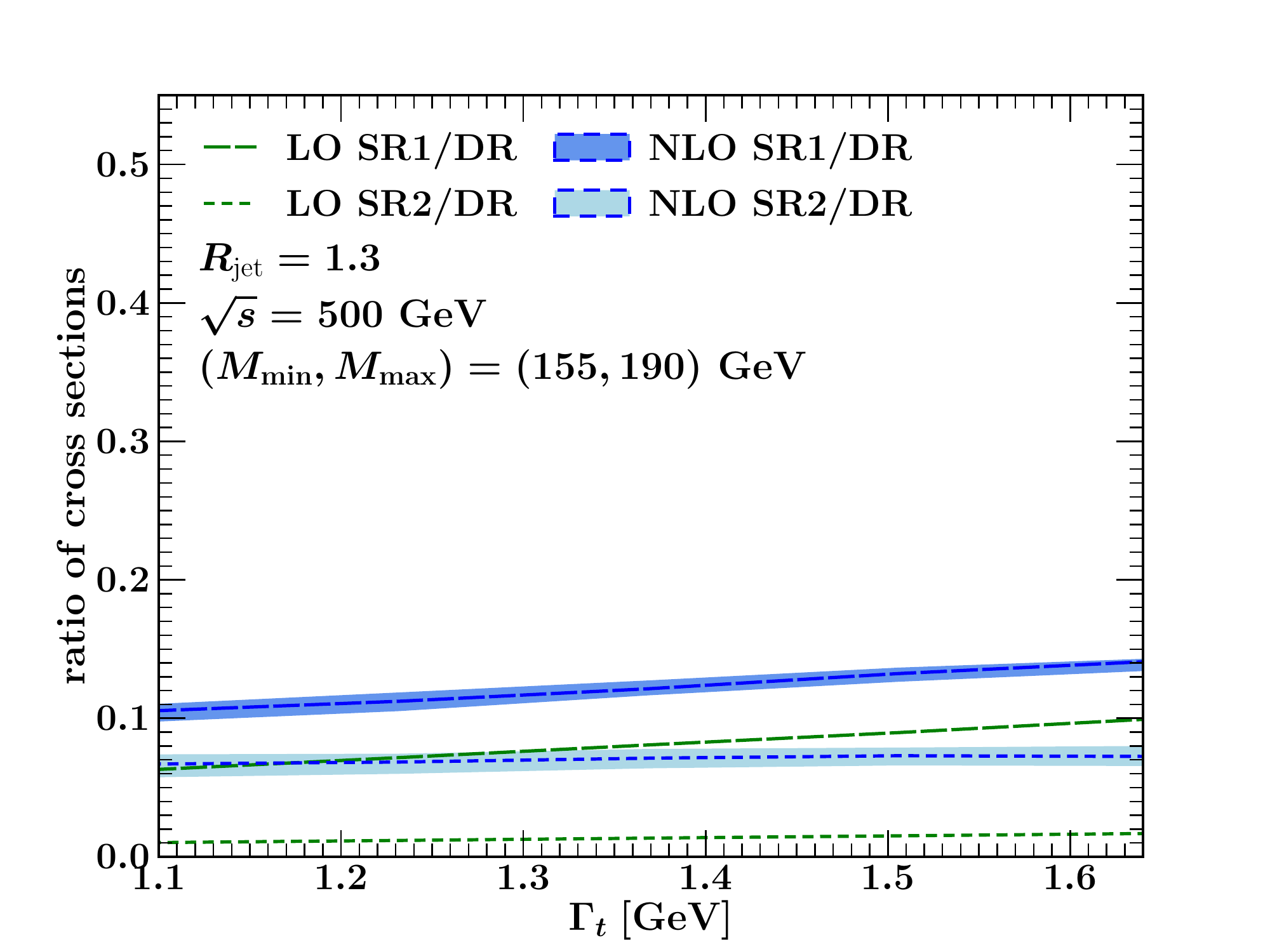} \\[-5pt]
(e) & (f)
\end{tabular}
\caption{Dependence on $\gamt$ of the ratios of the single resonant cross sections, SR1 (long dashed curves) and SR2 (short dashed curves) to the double resonant, DR, cross section, 
see \eq{xs-resonance-split} for definitions. Ratios of LO and NLO cross sections are shown in green and blue respectively and bands are obtained via variation
of $\muR$. 
The left-hand plots (a,c,e) show the results for $\rjet=0.5$ whilst the right-hand plots (b,d,f) illustrate the results for $\rjet=1.3$.
Plots (a,b), (c,d) and (e,f) illustrate the ratios for $(M_{\text{min}},M_{\text{max}}) = (165,180)$~GeV, $(160,185)$~GeV and $(155,190)$~GeV respectively.   
All results are for $\sqrt{s}=500$~GeV.
}
\label{fig:xs-structure-rjet}
\end{figure}

In each of the plots in \fig{xs-structure-rjet} the ratios $\sigma^{\text{SR1}}/\sigma^{\text{DR}}$ (long dashes) and 
$\sigma^{\text{SR2}}/\sigma^{\text{DR}}$ (short dashes)
are shown. LO and NLO results are in green and blue respectively and the bands around the NLO results indicate the 
uncertainty due to $\muR$-variation. 
From left-to-right in \fig{xs-structure-rjet} the jet radius parameter is varied, $\rjet \in \{0.5,1.3 \}$,
i.e. going left-to-right illustrates the effect of increasing the size of the jet-radius. 
From top-to-bottom in \fig{xs-structure-rjet} we have varied the definition of the resonance region -- specifically
a reconstructed top is defined to be resonant if $M(W^+,J_b) \in \{ (165,180), \; (160,185), \; (155,190)  \} $~GeV, i.e.
going top-to-bottom the resonance region is widened.  

In general, it is observed that at LO and NLO both ratios $\sigma^{\text{SR1}}/\sigma^{\text{DR}}$ and 
$\sigma^{\text{SR2}}/\sigma^{\text{DR}}$ display a roughly 
linear dependence on $\gamt$, as might be expected from the naive counting arguments given in 
the discussion preceding \eq{ratio-scaling}. 
For the latter ratio however the dependence is actually much flatter, whereas the former ratio 
importantly shows a stronger dependence on the top-quark width (i.e. the gradient is steeper).

For $\rjet=0.5$ at NLO, see plots (a), (c) and (e) of \fig{xs-structure-rjet}, 
the ratio $\sigma^{\text{SR2}}/\sigma^{\text{DR}}$ suffers from a large $\muR$-variation 
uncertainty (of real emission origin as discussed previously).
In contrast and as expected, $\sigma^{\text{SR1}}/\sigma^{\text{DR}}$ only has a small corresponding uncertainty for $\rjet=0.5$.
Widening the jet radius leads to the scale uncertainty increasing for $\sigma^{\text{SR1}}/\sigma^{\text{DR}}$ 
and decreasing for $\sigma^{\text{SR2}}/\sigma^{\text{DR}}$. 
For the results with $\rjet=1.3$ displayed in plots (b), (d) and (f) of \fig{xs-structure-rjet}, we see that the uncertainty 
bands of the two ratios have roughly the same thickness. 
As explained in \subsec{distr-mwb}, the reason for this is that a larger jet radius means that less radiation is leaked out of 
the $b$-jet for emissions from the 
top decay whilst unfortunately allows more radiation into the $b$-jet when emissions come from elsewhere in the $W^+W^-b\bar{b}$ 
process. These two effects combine to increase the size of NLO corrections and scale dependence of $M(W^+,J_b)< \mt$ and have 
the opposite effect for $M(W^+,J_b)> \mt$, as we also saw in \fig{mwb}.
The same reason lies behind the observed pattern that with increasing jet radius the ratios $\sigma^{\text{SR1}}/\sigma^{\text{DR}}$
and $\sigma^{\text{SR2}}/\sigma^{\text{DR}}$ are enhanced and diminished respectively.
By minimising non-top-resonance-decay radiation ending up in the $b$-jets, it is highly plausible that the use of modern 
jet-substructure techniques may help to control the behaviour of the SR1 region for increasing $\rjet$. 

As stressed above, precisely where one chooses to split the cross section into its various resonance regions 
is a little arbitrary and variations of this choice should be studied. In \fig{xs-structure-rjet} going from
top-to-bottom the ratios of cross sections decrease in size, however the pattern for the ratios of cross sections 
remains the same. This is due to the fact that by widening the definition of the resonances one 
naturally reduces the single resonant regions while at the same time increasing the double resonant one. 
The patterns remain the same because we have chosen to widen the resonance window in a symmetric manner.
Of course, one is free to pick different (e.g. asymmetric) resonance regions, however the patterns and results
we present here are unlikely to change dramatically. 

Overall the ratio $\sigma^{\text{SR1}}/\sigma^{\text{DR}}$ exploiting the region $M(W^+,J_b)>\mt$ appears as the more useful 
of the two to probe $\gamt$. 
The large scale dependence together with the flatness of the ratio $\sigma^{\text{SR2}}/\sigma^{\text{DR}}$ make 
this ratio rather unsuitable for a width-extraction.\footnote{Exactly the same conclusions hold for a ratio involving 
the total single resonant region, namely $(\sigma^{\text{SR1}}+\sigma^{\text{SR2}})/\sigma^{\text{DR}}$.} 
To quantify the potential sensitivity this approach may achieve and to compare the different setups,
we provide possible accuracies on $\gamt$ in two scenarios. Firstly, we assume a measurement
of the ratio of $\sigma^{\text{SR1}}/\sigma^{\text{DR}}$ with infinite experimental precision, which
due to our theoretical (scale) uncertainty translates into an uncertainty $\Delta\gamt^{\rm theo}$.
Secondly, we assume a fixed experimental error on the ratio $\sigma^{\text{SR1}}/\sigma^{\text{DR}}$ 
of $\pm 0.005$, which corresponds to a measured accuracy of the ratio of $5-10$\%. 
This enlarges the uncertainty of $\Delta\gamt^{\rm theo}$ to $\Delta\gamt$ as shown in \tab{sensitivity}. 
The table also depicts the setups presented in \sec{improvedwidth}. We conclude that better sensitivities 
are obtained for a small jet radius $\rjet$ in case of an unpolarised initial state and $\sqrt{s}=500$\,GeV.
We also observe that while increasing the interval $(M_{\text{min}},M_{\text{max}})$ generally 
improves the sensitivity $\Delta\gamt^{\rm theo}$, the actual number of events is 
diminished in the single resonant region leading to a smaller value for the ratio 
$\sigma^{\text{SR1}}/\sigma^{\text{DR}}$. 
When assuming an absolute error on the measurement of the ratio, this smaller value results in 
larger uncertainties  $\Delta\gamt$. 

We have checked that for $\sqrt{s}=500$\,GeV and an integrated luminosity of $500$\,fb$^{-1}$ several thousands of 
events can be recorded, even in the single resonant region SR1. The difficulties of the method are thus not in 
collecting enough statistics, but in a proper reconstruction of the invariant mass. 

\begin{table}
\begin{center}
\begin{tabular}{c  c  c  c  c  c}
\hline
$\sqrt{s}$\,[GeV] & Pol. & $\rjet$ & $(M_{\text{min}},M_{\text{max}})$\,[GeV] & $\Delta\gamt^{\rm theo}$\,[GeV] & $\Delta\gamt$\,[GeV] \\
\hline 
$500$ & unpol.    & $0.5$ & $(165,180)$ & $0.11$ & $0.19$\\
$500$ & unpol.    & $1.3$ & $(165,180)$ & $0.19$ & $0.27$\\ 
$500$ & unpol.    & $0.5$ & $(160,185)$ & $0.10$ & $0.20$\\
$500$ & unpol.    & $1.3$ & $(160,185)$ & $0.19$ & $0.31$\\
$500$ & unpol.    & $0.5$ & $(155,190)$ & $0.09$ & $0.25$\\
$500$ & unpol.    & $1.3$ & $(155,190)$ & $0.20$ & $0.36$\\
$500$ & $(-1,+1)$ & $0.5$ & $(165,180)$ & $0.07$ & $0.14$\\
$600$ & unpol.    & $0.5$ & $(165,180)$ & $0.12$ & $0.18$\\\hline
\end{tabular}
\end{center}
\vspace{-0.5cm}
\caption{Sensitivities on the top-quark width for different setups of the centre-of-mass energy $\sqrt{s}$,
the polarisation, given as $(P_{e^+},P_{e^-})$, the jet radius $\rjet$ and the 
interval $(M_{\text{min}},M_{\text{max}})$.}
\label{tab:sensitivity}
\end{table}

\subsection{Improved width extractions} 
\label{sec:improvedwidth}

There are a number of ways to exploit possible linear collider setups to improve the sensitivity of the 
method explored in the previous section on the top-quark width. 
The two ways we consider here are using polarised beams and increasing the centre of mass energy, both of 
which tend to enhance the proportion of single
resonant to double resonant contributions to the $W^+W^-b\bar{b}$ cross section. 

\subsubsection{Exploiting polarised beams}
\label{subsec:polar}

So far, our discussion has been based on simulations where the helicities of the incoming electron and positron were
averaged over.
However, a powerful feature of a linear collider is the fact that the initial state electron and positron beams
can be polarised. Given the electroweak nature of the primary interactions of the processes under consideration,
the inclusive cross section $\sigma$ can be decomposed according to~\cite{MoortgatPick:2005cw}
\begin{align}
\sigma = 
\frac{1}{4}\left(1-P_{e^+}\right)\left(1+P_{e^-}\right)\sigma_{-1+1}
+\frac{1}{4}\left(1+P_{e^+}\right)\left(1-P_{e^-}\right)\sigma_{+1-1}\,,
\label{eq:polarisation}
\end{align}
where $P_{e^+}$ and $P_{e^-}$ denote the relative polarisation of the positron and electron beam respectively.
$\sigma_{xy}$ encodes the cross section obtained with fixed helicities $x$ for the positron~$e^+$
and $y$ for the electron~$e^-$.
Whereas the double resonant diagrams contribute to both parts $\sigma_{-1+1}$ and $\sigma_{+1-1}$
several single resonant and non resonant diagrams
only contribute to the combination $\sigma_{+1-1}$.
Therefore, the single resonant contributions can be enhanced by choosing the $(P_{e^+},P_{e^-})=(+1,-1)$ 
combination.
We note that \eq{polarisation} is also valid at the level of differential cross sections, i.e.
we can replace all occurrences of the inclusive cross section $\sigma$ with e.g. $d\sigma/dM(W^+,J_b)$
at LO and NLO in QCD. We obtained polarised initial states in \mga by adapting the model
files such as to select only left- or right-handed couplings appropriately. 
We have validated our results at LO through a 
comparison to results with explicit polarisations (available in the LO version of the \mga code) and at 
NLO by ensuring that we could reproduce the unpolarised cross section using \eq{polarisation}.

\begin{figure}[h]
\begin{tabular}{c c}
\hspace{-1.0cm} \includegraphics[trim=0cm 0cm 0cm 1.3cm,clip,width=8.4cm]{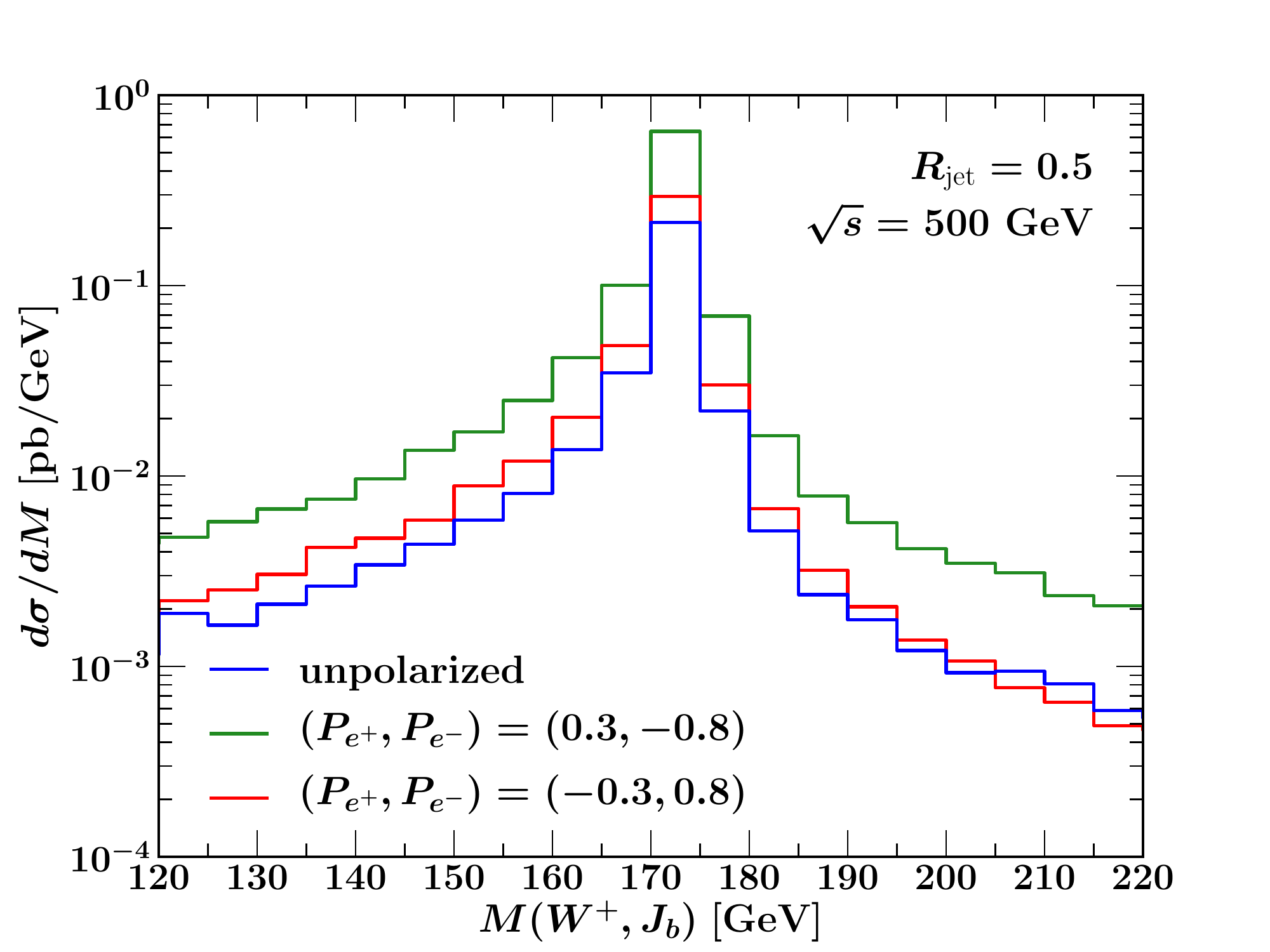}  
& \hspace{-1.2cm} \includegraphics[trim=0cm 0cm 0cm 1.3cm,clip,width=8.4cm]{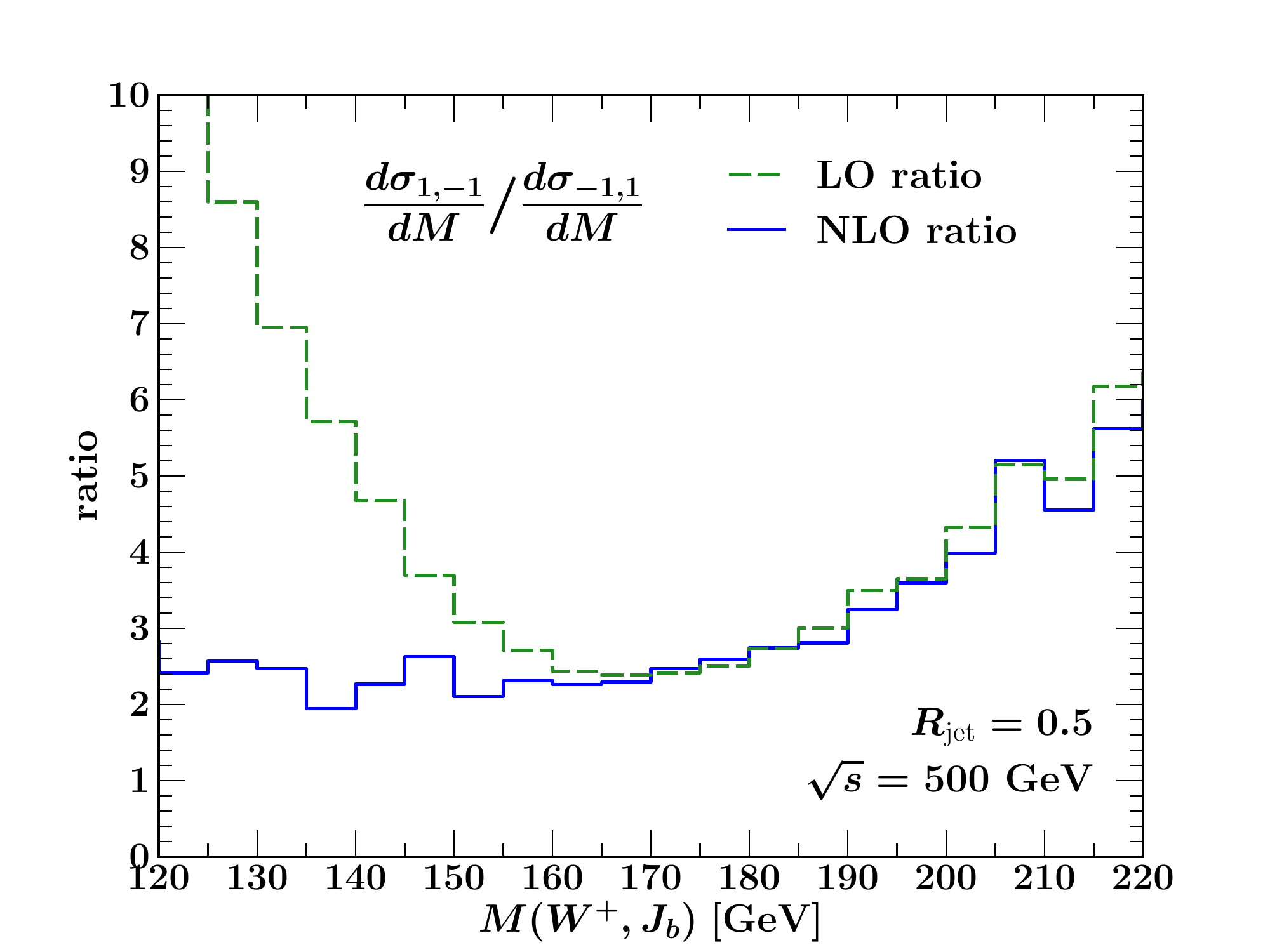} \\[-5pt]
(a) & (b)
\end{tabular}
\caption{(a) NLO cross section as a function of $M(W^+,J_b)$ in GeV for different polarisations of the initial 
electron and positron: unpolarised (blue), $(P_{e^+},P_{e^-})=(0.3,-0.8)$ (green) and 
$(P_{e^+},P_{e^-})=(-0.3,+0.8)$ (red);
(b) Ratio of cross sections with $(P_{e^+},P_{e^-})=(1,-1)$ over $(P_{e^+},P_{e^-})=(-1,1)$
at LO (dashed green) and NLO (black) as a function of $M(W^+,J_b)$ in GeV.
All results are for $\sqrt{s}=500$~GeV and $\rjet=0.5$.
We note that the results for $\rjet=1.3$ (not shown) are similar.}
\label{fig:XSpol}
\end{figure}

In \fig{XSpol}(a) we show the differential cross section as a function of $M(W^+,J_b)$ for the
unpolarised initial state as discussed beforehand, but also for two common polarisations
$(P_{e^+},P_{e^-})=(0.3,-0.8)$ and $(P_{e^+},P_{e^-})=(-0.3,+0.8)$ at a linear collider.\footnote{The
chosen polarisation degrees are those foreseen for the current baseline design,
however, higher polarisation degrees for both beams could be achieved at a later stage.}
We find that the inclusive cross section increases the closer the polarisation is to
$\sigma_{+1-1}$, i.e. $(P_{e^+},P_{e^-})=(+1,-1)$. 
This increase is even more pronounced for the single resonant part of the cross section, 
as is apparent in \fig{XSpol}(b).
In the latter figure we show the ratio of the combination
$\sigma_{-1+1}$ over the combination $\sigma_{+1-1}$,
both at LO and NLO QCD. Whereas at LO the rise for low values of
$M(W^+,J_b)$ is dramatic due to the vanishing of single and non resonant diagrams
for $\sigma_{-1+1}$, the ratio at NLO QCD remains rather constant in
the region $M(W^+,J_b)\leq \mt$.
Again this effect is induced by the real emission of gluons from double resonant diagrams.
As expected LO and NLO ratios
are much closer in the single resonant region SR1 with $M(W^+,J_b)> \mt$.

\begin{figure}
\begin{tabular}{c c}
\hspace{-1.0cm} \includegraphics[trim=0cm 0cm 0cm 1.3cm,clip,width=8.4cm]{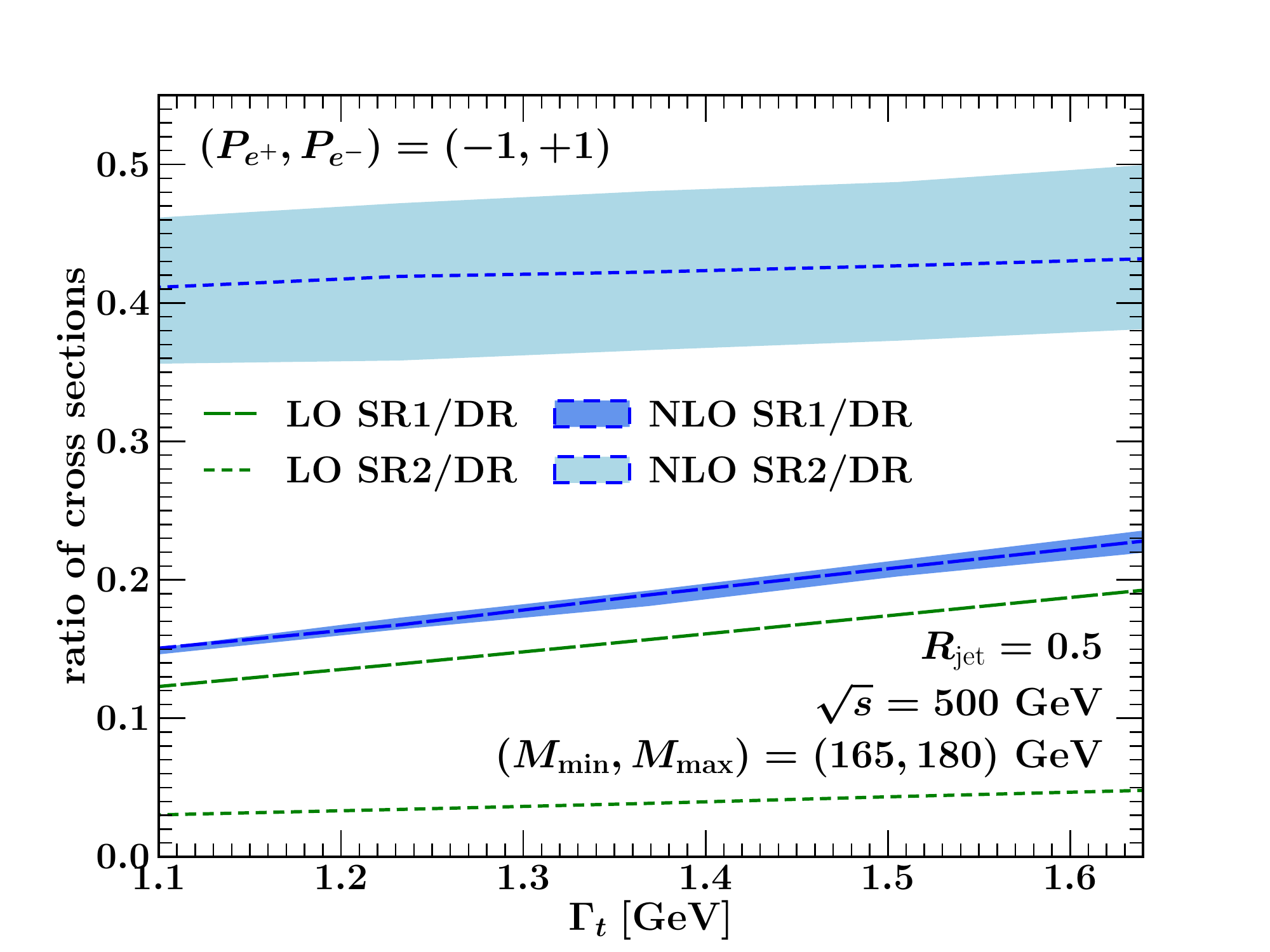}  
& \hspace{-1.2cm} \includegraphics[trim=0cm 0cm 0cm 1.3cm,clip,width=8.4cm]{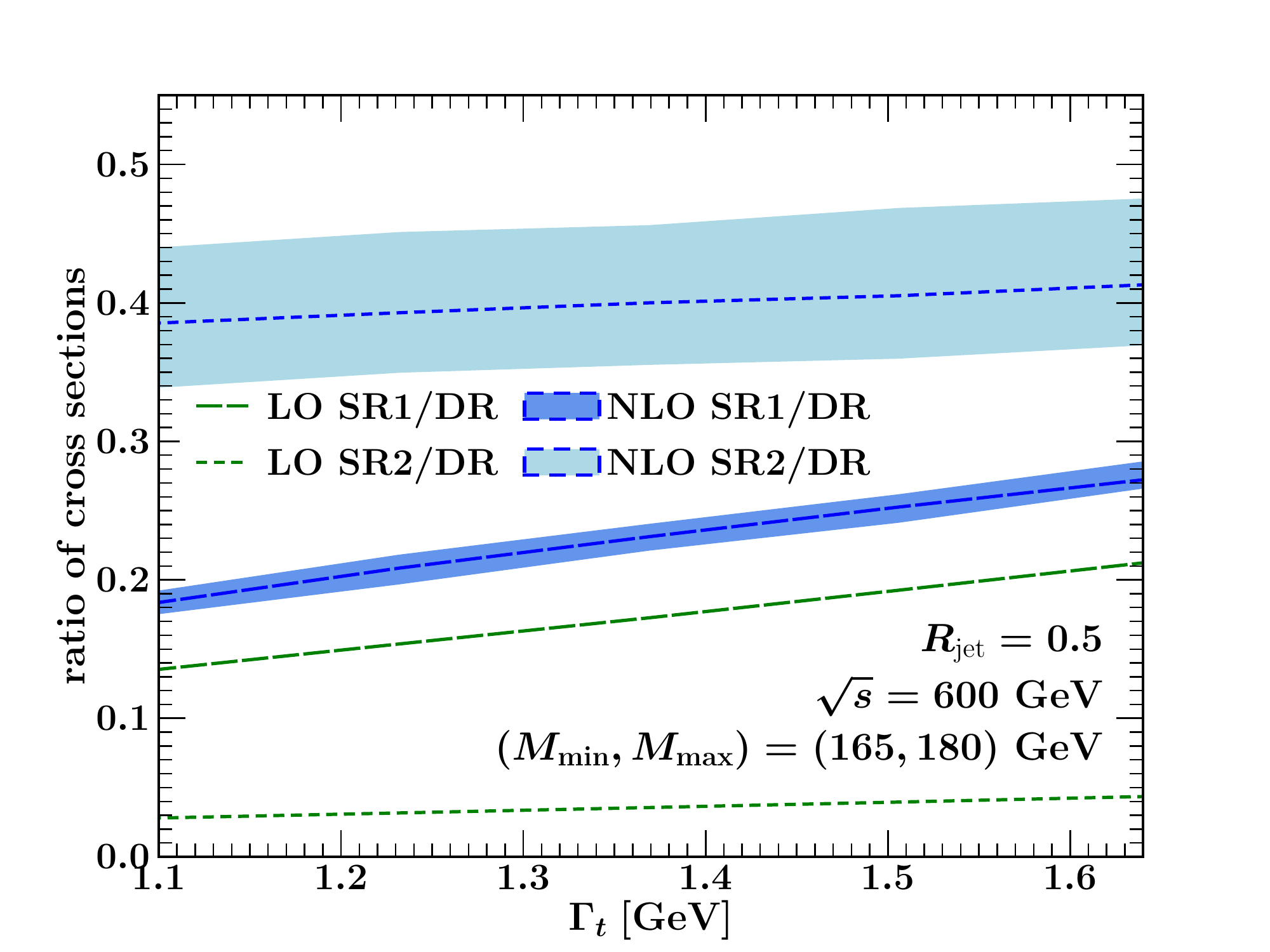} \\[-5pt] 
(a) & (b)
\end{tabular}
\caption{Dependence on $\gamt$ of the ratios of the single resonant cross sections, SR1 (long dashed curves) and SR2 (short dashed curves) to the double resonant, DR, cross section, 
see \eq{xs-resonance-split} for definitions. Ratios of LO and NLO cross sections are shown in green and blue respectively and bands are obtained via variation
of $\muR$. 
Plot (a) shows the results for polarised beams $(P_{e^+},P_{e^-}) = (-1,+1)$ at $\sqrt{s}=500$~GeV, whilst plot (b)
shows the patterns for unpolarised beams for a higher centre of mass energy of $\sqrt{s}=600$~GeV. 
All results are for $\rjet=0.5$.}
\label{fig:gamt-variation-structure-pol}
\end{figure}

Taking all of this into account, it is evident that the sensitivity to the top-quark width 
can thus be significantly increased by the combination $(P_{e^+},P_{e^-})=(-1,+1)$. 
For this particular case we show the corresponding sensitivity in \fig{gamt-variation-structure-pol}
obtained for a jet radius of $R_{\text{jet}}=0.5$.
Compared to the unpolarised case (\fig{xs-structure-rjet}(a)) we see that not only is the 
ratio $\sigma^{\text{SR1}}/\sigma^{\text{DR}}$ increased in size,
but also its gradient is visibly enhanced (by about $23$\,\% for the setup considered). 
We perform the study of extracting $\gamt$ in the two scenarios described at the end of \subsec{xs-structure}
and indicate the corresponding accuracies in \tab{sensitivity}, which we find to be significantly 
improved compared to the unpolarised case.

\subsubsection{Exploiting a higher centre of mass energy} \label{subsec:com-energy}

Due to the increase by more than a factor of $4$ of the cross section of $e^+e^-\rightarrow t\bar t H$
(relevant for the measurement of the top-Yukawa coupling) when going from a centre of mass energy of $500$\,GeV to $600$\,GeV, 
a slightly higher initial centre of mass energy at a linear collider is well-motivated.
Hence, in this subsection we present the sensitivity on the top-quark width for $\sqrt{s}=600$\,GeV. 
Once again we study the ratios of \eq{xs-ratios}, showing these in \fig{gamt-variation-structure-pol}(b),
and again using $\rjet=0.5$.
We again find that not only is the relevant ratio $\sigma^{\text{SR1}}/\sigma^{\text{DR}}$ increased in size, 
but additionally the sensitivity on the top-quark width is enhanced, with the gradient of the slope increased 
by about $38$\,\% in the setup considered.
Even though the scale uncertainty increases in size (thus increasing $\Delta\gamt^{\text{theo}}$), 
the accuracy on $\gamt$ when an experimental error is included, $\Delta\gamt$, is still slightly improved
compared to the case of $\sqrt{s}=500$~GeV due to the enhancement of the gradient. 
The relevant numbers on the sensitivity can be found in \tab{sensitivity}.

\subsection{Opportunities and limitations of the method} \label{subsec:limitations}

We have shown that the ratio $\sigma^{\text{SR1}}/\sigma^{\text{DR}}$ is a promising observable for extracting 
the top-quark width, independently from $g_{tbW}$, in a generic analysis of the $W^+W^-b\bar{b}$ 
process at a linear collider. 
An extraction of $\gamt$ for different choices of the resonance windows used (\eq{xs-resonance-split}) 
not only provides an in-built consistency check on the method and measurements, but also (through 
the combination of these extractions) may allow for the shrinking of uncertainties in $\gamt$.
Furthermore, we emphasise that providing a method complementary to a lineshape fit, that 
additionally allows for deviations of $\gamt$ independently of variations in $g_{tbW}$, is of significant 
value. An interesting avenue to explore would also be to assume a fixed value for $\gamt$ and 
investigate the extent to which our method can disentangle the $g_{ttV}$ couplings 
from $g_{tbW}$ (top-pair production is sensitive to the product $g_{ttV}g_{tbW}^2$).

For the investigation we have presented in this work some assumptions have been made. Firstly, we have 
assumed a perfect $b$ and $\bar{b}$ jet-tagging as well as a perfect reconstruction of $W$-bosons. A
more sophisticated analysis could include errors due to mistagging etc. 
Such uncertainties are unlikely to affect the theory results strongly and can rather be 
included as an experimental error. 
An additional assumption we have made is that the couplings appearing in the amplitudes for 
$e^+e^- \to W^+W^-b\bar{b}$ all take their (fixed) \sm{} values. Clearly if couplings such as $g_{ttZ}$ 
were to differ from their \sm{} value, then the ratios predicted would also change, thus skewing the extracted width.  
This potential problem can be overcome by using as inputs to our method, values for the couplings 
as constrained in other collider processes. 

The simulation underlying this work is a parton-level simulation, namely one that does not include 
the effects of parton-showering and hadronization. These two steps beyond a fixed-order simulation
are known to alter some distribution shapes significantly. While it is therefore important to 
extend our results to include these effects,\footnote{This can be
done within the framework of \mga as well as that of the \whz{}, though
the consistent matching to parton shower of the $W^+W^-b\bar{b}$ process is not totally 
straightforward and requires care due to the presence 
of intermediate coloured resonances (see the discussions in \mycites{Campbell:2014kua,Jezo:2015aia}). }  
(and thus any potential shape distortions to invariant mass distributions)
we expect that, after parton-showering and hadronization, the changes to the resonance regions 
that arise from variations in $\gamt$ are very similar to those observed in our fixed-order analysis,
and therefore that the ratios remain a very good probe of $\gamt$.
Moreover, jet-substructure techniques could be employed to understand and control radiation in 
an event such that the split into the resonance regions and the structure of the cross section
within these is not altered significantly from the fixed-order analysis we have discussed.
Therefore, we fully expect that our conclusions and the usefulness of the method to be largely
unaltered. 

We end this section with a few comments regarding the applicability of this method for $\gamt$-extraction at the LHC from 
the $p \, p \to W^+W^-b\bar{b}$ process. This process has received significant attention recently and NLO QCD
corrections to the full process are known \cite{Bevilacqua:2010qb,Denner:2010jp,Denner:2012yc,Frederix:2013gra,Cascioli:2013wga}.
While certainly a possibility worth exploring (one which is however beyond the scope of this work) the proton-proton 
initiated process intrinsically contains some difficulties.
The ratio of the leading parts of the squared matrix element in the single and double resonant regions will in principle 
be sensitive to $\gamt$, for the same arguments presented in \sec{offshell-width}.
However, given that at LO the squared matrix element is proportional to $\as^2$ and that the predicted cross section 
additionally carries a dependence on a factorization scale, $\muF$, the uncertainty due to the variation of these scales 
is significantly larger than that observed in this study (see discussions in the references cited above), in particular
for exclusive observables. Though the ratio $\sigma^{\text{SR1}}/\sigma^{\text{DR}}$ may indeed be quite sensitive 
to $\gamt$, we feel it is very likely that the uncertainty on the ratio would make an extraction prohibitive in 
practice (much like we have demonstrated for the ratio $\sigma^{\text{SR2}}/\sigma^{\text{DR}}$), 
even using the state-of-the-art NLO computations. It is of course possible, that with some modifications or in combination
with additional measurements, such ratios would also be useful in a hadron-collider environment.  

%%%%%%%%%%%%%%%%%%%%%%%%%%%%%%%%%%%%%%%%%%%%%%%%%%%%%%%%%%%%%%%%%%%%%%%%%%%%%%%%%%%%%
\section{Conclusions} \label{sec:conclusion}
%%%%%%%%%%%%%%%%%%%%%%%%%%%%%%%%%%%%%%%%%%%%%%%%%%%%%%%%%%%%%%%%%%%%%%%%%%%%%%%%%%%%%

We have performed a detailed study of the $e^+e^-\to W^+W^- b\bar{b}$ process at NLO in QCD using \mga to 
simulate the fixed-order results. In particular we have examined the structure of reconstructed top-quark
masses which has allowed for a detailed understanding of the double, single and non resonant contributions of the 
total cross section. We have used this to show that the ratio of single resonant to 
double resonant cross section contributions is sensitive to the top-quark width whilst simultaneously
being independent of the $g_{tbW}$ coupling.   
The central results of this article are the in-depth investigation of this ratio. 
We have shown in a typical linear collider analysis, that with a careful definition or choice of the single 
resonant region of the cross section, that such a ratio is, also in practice, sensitive to the value of $\gamt$, and can be 
exploited to extract the width at an $e^+e^-$ collider. We have explored the effects that variations
in both the jet radius as well as in the resonance window (in which reconstructed top quarks are defined to be resonant)
have on the ratios. 
Additionally, we showed that using polarised beams or higher centre of mass energies leads to
an enhanced sensitivity to $\gamt$.
In a study of the expected errors in the extraction of $\gamt$ using this method, we find that 
attainable accuracies of $< 200$~MeV are possible with unpolarised beams at $\sqrt{s}=500$~GeV. 
We note that these are comparable to the accuracies quoted in the literature obtained from 
invariant-mass lineshape fitting, and that they can be significantly improved by exploiting polarised 
initial states. 

Our study of the $e^+e^-\to W^+W^- b\bar{b}$ process here has been restricted to fixed-order. 
The next step in extending this investigation is to include effects due to parton-showering and hadronization, 
which we look forward to investigating in future work. 

%%%%%%%%%%%%%%%%%%%%%%%%%%%%%%%%%%%%%%%%%%%%%%%%%%%%%%%%%%%%%%%%%%%%%%%%%%%%%%%%%%%%%
\section{Acknowledgements} 
This work has received the support of the Collaborative Research Center SFB676 of the DFG, ``Particles, Strings, and the Early Universe."
The work of AP is supported by the UK Science and Technology Facilities Council [grant ST/L002760/1]. 
%%%%%%%%%%%%%%%%%%%%%%%%%%%%%%%%%%%%%%%%%%%%%%%%%%%%%%%%%%%%%%%%%%%%%%%%%%%%%%%%%%%%%

%\newpage
\phantomsection
\addcontentsline{toc}{section}{References}
\bibliographystyle{JHEP}
\bibliography{ilc_width_refs.bib}

\end{document}